\documentclass[useAMS, usenatbib]{mn2e}

\usepackage{graphicx}
\usepackage{mathptmx}
\usepackage{supertabular}

\newcommand{\ciii}{C~\textsc{iii}}
\newcommand{\civ}{C~\textsc{iv}}

\newcommand{\oiv}{O~\textsc{iv}]}

\newcommand{\ovi}{O~\textsc{vi}}

\newcommand{\siiv}{Si~\textsc{iv}}
\newcommand{\nv}{N~\textsc{v}}
\newcommand{\niii}{N~\textsc{iii}}
\newcommand{\hi}{H~\textsc{i}}

\newcommand{\lyalpha}{Ly $\alpha$ }
\newcommand{\kms}{km~s$^{-1}$}
\newcommand{\cmN}{cm$^{-2}$}

\newcommand{\lam}{$\lambda$}
\newcommand{\aj}{AJ} 
\newcommand{\mnras}{MNRAS} 
\newcommand{\apj}{ApJ} 
\newcommand{\apjl}{ApJ} 
\newcommand{\apjs}{ApJS} 
\newcommand{\aap}{A\&A} 
\newcommand{\aaps}{A\&AS} 
\newcommand{\araa}{ARA\&A} 

\newcommand{\pasp}{PASP} 

\newcommand{\nat}{Nat} 

\title[Narrow Absorption Line Complex at z = 3.45]{The Origins of a Rich Absorption Line Complex in a Quasar at Redshift 3.45}

\author[L. E. Simon and F. Hamann]{Leah E. Simon$^{1}$\thanks{email: {\tt lsimon@astro.ufl.edu} (LES); {\tt hamann@astro.ufl.edu} (FH)} and Fred Hamann$^{1}$\\
$^{1}$Department of Astronomy, University of Florida, 211 Bryant Space Science Center, Gainesville, FL 32611, USA }

\begin{document}

\maketitle

\begin{abstract}

We discuss the nature and origin of a rich complex of narrow absorption lines in the quasar J102325.31+514251.0 at redshift 3.447. We measure nine \civ(\lam1548,1551) absorption line systems with velocities from --1400 to --6200~\kms, and full widths at half minimum ranging from 16 to 350~\kms.  We also detect other absorption lines in these systems, including \hi, \ciii, \nv, \ovi, and \siiv .  Lower ionisation lines are not present, indicating a generally high degree of ionisation in all nine systems.  The total hydrogen column densities range from $\la$$10^{17.2}$ to $10^{19.1}$~cm$^{-2}$.  The tight grouping of these lines in the quasar spectrum suggests that most or all of the absorbing regions are
physically related. We examine several diagnostics to estimate more directly the location and origin of each absorber. Four of the systems can be attributed to a quasar-driven outflow based on line profiles that are smooth and broad compared to thermal line widths and to the typical absorption lines formed in intergalactic gas or galaxy halos.  Several systems also have other indicators of a quasar outflow origin, including partial covering of the quasar emission source (e.g., in systems with speeds too high for a starburst-driven flow), \ovi\ column densities above $10^{15}$ cm$^{-2}$ and an apparent line-lock in \civ\ (in two of the narrow profile systems). A search for line variability yielded null results, although with very poor constraints because the comparison spectra have much lower resolution. Altogether (but not including the tentative line-lock) there is direct evidence for 6 of the 9 systems forming in a quasar outflow. Consistent with a near-quasar origin, eight of the systems have metallicity values or lower limits in the range $Z \ge 1-8~Z_{\odot}$. The lowest velocity system, which has an ambiguous location based on the diagnostics mentioned above, also has the lowest metallicity, $Z\leq 0.3~Z_{\odot}$, and might form in a non-outflow environment farther from the quasar. Overall, however, this complex of narrow absorption lines can be identified with a highly structured, multi-component outflow from the quasar.  The high metallicities are similar to those derived for other quasars at similar redshifts and luminosities, and are consistent with evolution
scenarios wherein quasars appear after the main episodes of star formation and metal enrichment in the host galaxies.

\end{abstract}


\section{Introduction}
\label{sec:intro}

Quasars represent episodes of rapid  supermassive black hole (SMBH) growth and probably a unique period in the early evolution of galaxies.  They may directly follow a major merger \citep{PerezGonzalez08, Hopkins08} or a big blowout of gas and dust.  However, the nature of the relationship between SMBH growth and galaxy formation is not well understood.  Feedback from quasar outflows may play an important role in the evolution of this relationship.  We are using narrow absorption lines (NALs) in quasar spectra to study quasar outflows and environments across a range of scales.

NALs have full widths at half minimum (FWHMs) less than several hundred~\kms, and they appear in a variety of ultraviolet (UV) resonance transitions, including \civ~\lam1548 and~\lam1551, \nv~\lam1239 and~\lam1243, \ciii~\lam977, \siiv~\lam1394 and~\lam1402 and \lyalpha~\lam1216, Ly$\beta$~\lam1026, Ly$\gamma$~\lam973, Ly$\delta$~\lam950, and Ly$\epsilon$~\lam938 \citep{Foltz86, Anderson87, Hamann04}.  The first step in using these absorption lines to study the quasar environment is to determine simply where the lines form. There are several possibilities, including unrelated (intervening) clouds and galaxies which happen to lie in the line-of-sight, nearby cluster galaxies and intrinsic clouds within the quasar host galaxy and its extended environment \citep{Ganguly01, Vestergaard03, Trump06, Nestor08, Wild08, Gibson08}.  

The statistical excess of \civ\ NALs at velocity shifts v~$> -12,000$~\kms, where negative v indicates motion towards the observer, indicates that many of these absorbers are directly related to quasar environments \citep{Weymann79, Nestor08, Wild08}. The excess is largest at v~$\ga -1000$~\kms, where roughly 80\% of \civ\ systems with rest equivalent width \textsc{REW}(1548\AA)~$\geq 0.3$~\AA\ have a quasar--related origin \citep{Nestor08}.  This intrinsic gas can form in quasar--driven outflows, starburst--driven outflows, merger remnants or ambient gas in the host halos, or in other galaxy halos in the same galaxy cluster as the quasar.  At higher velocities, the excess can be attributed directly to quasar--driven outflows.  In the velocity range -1000 to -12,000~\kms, \citet{Nestor08} estimate that $\ga$~43\% of \civ\ NALs with \textsc{REW}(1548\AA)~$\geq$~0.3~\AA\ originate in a quasar outflow.

NALs encompass a wealth of information about the basic properties of quasar outflows, which we use to gain insights into the outflow physics, acceleration mechanisms and geometry of the near quasar environment.  NALs represent a very different type of quasar outflow compared to the well-studied much broader and higher velocity broad absorption lines (BALs), but they might simply be different manifestations of a single outflow phenomenon viewed at different angles \citep{Ganguly99, Ganguly01, Elvis00}. The NALs that do not form in outflows probe the gaseous environments of quasars more generally, and can be used to examine conditions in different host galaxies environments such as  starburst--driven outflows or larger-scale gas distributions produced by messy mergers.

We are involved in a program to study the location, origin and abundance information for absorbers in a sample of high redshift quasars.  We are particularly interested in high redshift quasars because z~$\sim 2-4$ is the cosmic era when host galaxies are thought to grow rapidly and form most of their stars, possibly through merger events \citep{PerezGonzalez08, Hopkins08}.  Choosing redshifts above z~$\sim$~2.7--3 also allows us to measure lines at shorter rest-frame wavelengths with ground--based telescopes, including importantly, the \hi\ Lyman series, in order to obtain more and better constraints on the absorber ionisations, column densities and metal abundance.  We are interested in using abundances to discern rough star formation histories of quasar host galaxies in order to make inferences about the relationship between the quasar, the growth of the central black hole, and the evolution of the host galaxy.

Broad emission lines (BELs) have been used most often to study quasar abundances. The most reliable results suggest metallicities of at least solar, and up to a few times solar, which requires significant previous star formation in the host \citep{Hamann99, Hamann02, Dietrich03, Warner04, Nagao06a, Simon10}. The metal-rich BEL result is true even for the highest redshifts studied, e.g. \citet{Pentericci02, Jiang07, Juarez09}, with redshifts out to 6.4.  The most reliable results based on BAL column densities suggest metallicity ranges between solar and ten times solar \citep{Arav01}.  Previous studies of NALs in low redshift samples have found super-solar metallicities and highly ionised gas, and have successfully probed several other NAL outflow characteristics \citep{Hamann99, Ganguly03, Dodorico04, Hutsemekers04, Ganguly06, Gabel06}.  These lower redshift samples cover a wide range of luminosities, observed in the ultraviolet (UV) spectral range, where many useful metal and Hydrogen lines occur.  

NALs offer certain advantages in the study of metallicities and other gas characteristics in the near-quasar environment.  Their narrow widths mean the \civ\ doublets, separated by 500~\kms, are resolved.  We use resolved absorption line doublets to disentangle saturation effects, and to obtain accurate line optical depth and column density measurements.  NALs also form in a range of physical locations, providing a more complete picture of the regions near quasars.  Because the NAL methods are completely independent from the BEL methods, requiring only column densities and ionisation corrections, the NAL metallicities provide an independent test of the BEL results. 

Here we present results for the particular luminous quasar J102325.31+514251.0 (hereafter J1023+5142) at a redshift of $z_{em} = 3.447$, which is during the peak in quasar activity and the epoch of rapid galaxy formation. This quasar contains a rich complex of nine distinct \civ\ NAL systems at velocities from -1400 to -6200~\kms.  The density and diversity of lines in this complex merits special attention.    We will argue below that most (or all) of these systems form in a highly structured quasar--driven outflow.  

To interprete the metallicities and other data provided by these NALs, we examine several diagnostics that can identify intrinsic NALs that form in quasar--driven outflows \citep{Hamann97, Barlow97}.  In particular,  1) variability studies have found intrinsic absorbers varying on relatively short timescales of months to years, providing strong evidence for these absorbers belonging to outflows either crossing the line of sight to the quasar or experiencing changing ionisation with the variations in the continuum emission \citep{Hamann97, Barlow97, Aldcroft97, Narayanan04, Misawa07c}.  2) Detection of partial coverage of the background light source along the line of sight strongly implies gas forming very near the source.  This phenomenon occurs when the absorbing 'clouds' are smaller than the background source, allowing part of the light from the source to reach the observer unabsorbed.  This partial covering is easily detected in multiplets like the \civ\ doublet where the optical depth ratio between the two lines is fixed by the oscillator strengths.  When the source is partially covered, some light fills in the bottom of the absorption line, and makes the apparent optical depth ratio appear different than the real optical depth ratio. 3) Outflow lines tend to have profiles that are broad and smooth compared to thermal widths \citep{Hamann99, Srianand00, Ganguly06, Schaye07}.  In well studied NALs, these three indicators (variability, partial covering and broad profiles) tend to appear together, which further increases the probability that the occurrence of an individual indicator accurately predicts an outflow origin very near the quasar for a given absorption line (see also \citet{Hamann08} and references therein).  We also note that super--solar metallicities are consistent with an intrinsic origin for the gas.  There are examples of high--metallicity gas in intervening systems, but not of low--metallicity intrinsic gas \citep{Prochaska06, Schaye07}. 

We describe the data acquisition and reduction in \S~\ref{sec:obs}, the identification and fitting of the absorption lines in \S~\ref{subsec:id} and \ref{subsec:fit} and the abundance and ionisation analysis in \S~\ref{sec:ion}.  We briefly describe individual absorption line systems in \S~\ref{sec:sys}. We discuss the arguments for the locations, probable intrinsic origins and quasar--driven outflow properties of the gas in \S~\ref{sec:disc} and conclude with a summary in \S~\ref{sec:sum}.  

\section{Observations and Data Reduction}
\label{sec:obs}

\begin{figure*}
\vspace{4mm}
\includegraphics{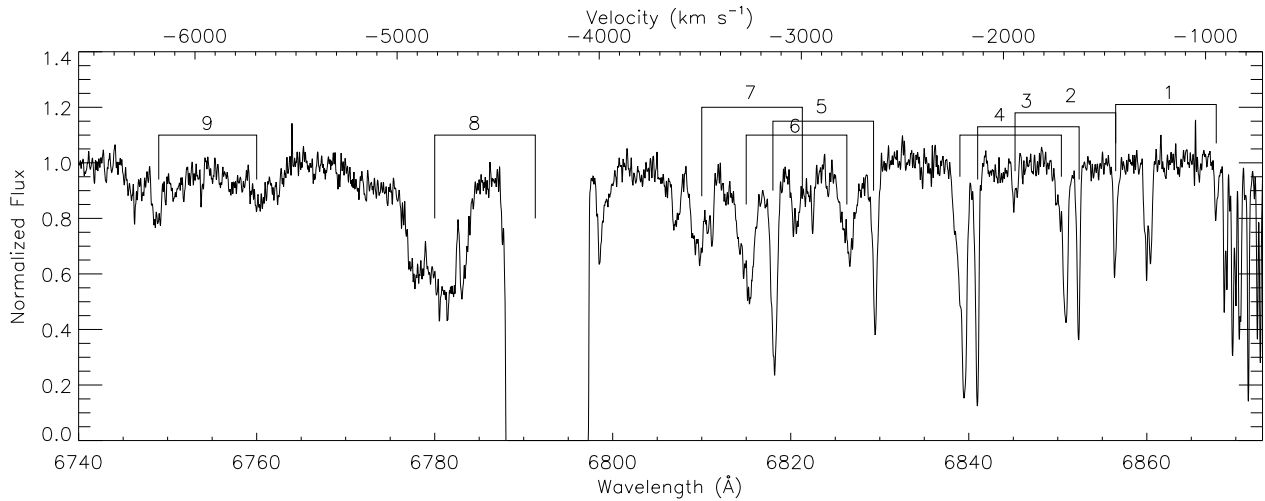}

\caption{Region of the spectrum of J1023+5142 with \civ\ absorption.
  Individual \civ\ doublets are labelled by number.  The lower x-axis is observed wavelength in Angstroms, while the upper x-axis is velocity shift of the shorter wavelength doublet line at 1548.20~\AA\ from the quasar rest frame in kilometres per second.  The flux units are normalised so that the continuum has a value of one.  The gap between 6788 and 6797~\AA\ is a gap between Echelle orders in the spectrograph.  The longer wavelength line of system 8 falls in this gap.  Strongly blended lines are considered components of a single system, e.g. systems 7 and 9.}
\label{fig:civ}
\end{figure*}

We observed the quasar J1023+5142 on March 29, 2007 with the Keck I
HIRESr Echelle spectrograph as part of an observing campaign to
measure spectra of high redshift quasars with known narrow associated
absorption lines.  We used an 0''.86 wide slit for a spectral resolution of R~$\sim 40,000$ or velocity resolution of $\sim 7$~\kms.  Our data span the wavelength range from 3700 to 8100~\AA\ corresponding to 830 to 1820 \AA\ in the quasar rest frame.  This spectral range covers a variety of interesting lines, including rest-frame Ly~$\gamma~970$~\AA, \civ~1548, 1551~\AA\ and several other lines in the \hi\ Lyman series down to the Lyman limit at 912~\AA.  We use four exposures totalling 2 hours on the source.  The spectral region from $\sim 3700$ to $\sim 4980$~\AA\ is well-covered with considerable overlap between Echelle orders at some wavelengths, but above 6540~\AA\ there are small 15-40~\AA\ gaps between orders (one such gap is apparent in Figure~\ref{fig:civ}), and two larger gaps at 4980-5070~\AA\ and 6575-6670~\AA\ where the spectrum falls into a physical gap between detectors.  

We reduce all the data using the \textsc{MAKEE HIRES} data reduction package. The spectra are sky background-subtracted and extracted from the 2D frame with a low order polynomial trace using a white dwarf standard observed the previous night with similar seeing conditions. We use a Thorium Argon (ThAr) lamp spectrum for wavelength calibration. The resulting spectra are on a vacuum and heliocentric wavelength scale. The spectra are not absolute flux calibrated.

\begin{figure}
\centering
\vspace{4mm}
\includegraphics[width=8cm]{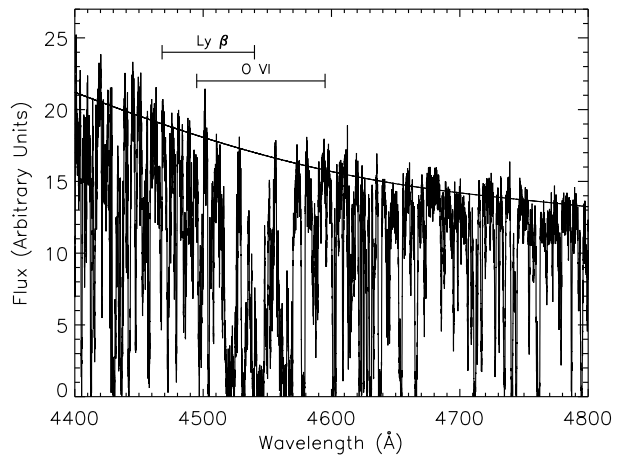}
\caption{Region of \lyalpha\ forest spectrum with the continuum fit over-plotted. The regions spanning the Ly~$\beta$ and \ovi~ NALs are labelled above the spectrum. }
\label{fig:contin}
\end{figure}

We normalise the spectra to unity by fitting a pseudo-continuum to all of the quasar emission, including the emission lines.  The pseudo-continuum is defined as follows: for regions with few absorption lines, we apply a polynomial fit to the local continuum in each Echelle order. We accomplish the continuum normalisation in crowded regions where the continuum is affected by significant absorption, e.g. the \lyalpha\ forest, by first averaging together several adjacent spectral orders into a single spectrum.  Then, we visually inspect the region for small sections of continuum not affected by absorption or obvious noise spikes, and interpolate between these sections, fitting the entire region with a low order polynomial.  Our continuum fit for a region of the \lyalpha forest containing the Ly$\beta$ and \ovi\ NALs is shown in Figure~\ref{fig:contin}.  The continuum placement has an uncertainty of $\sim 10$\% in the forest and 2-3\% at other wavelengths.

\section{Analysis}
\label{sec:anal}

\subsection{Identification}
\label{subsec:id}

The broad, flat shape of the emission features in the spectrum of J1023+5142 make an accurate emission redshift difficult to determine.  The redshift provided by the SDSS spectrum is $z_{em} = 3.447$.  We estimate the reliability of this value by measuring the redshifts of the \civ~\lam1549, \ciii]~\lam1909  and \siiv+\oiv~\lam1398 emission lines using measurements of their centroids.  We shift each centroid respectively by -824, -730 and +36~\kms\ to correct for known offsets from the nominal quasar redshift (\textsc{[O III]}~\lam5007 emission), based on measurements by \citet{Tytler92} and \citet{Shen07} for average quasars, where negative values are blueshifts.   The average redshift obtained from these emission lines is $z_{em}= 3.429$, which is offset from the SDSS value by $\Delta z = 0.018$ or $\sim -1200$~\kms.  These results and our efforts to measure the inherently uncertain emission line centroids suggest that the uncertainty of the redshift measured by SDSS is not more than  $\Delta z_{em} \leq 0.02$, corresponding to $\Delta \mbox{v} \leq -1350$~\kms.  We adopt the SDSS value throughout the remainder of this paper.

We identify nine distinct \civ\ absorption line systems within
6200~\kms\ of the quasar redshift. We will refer to these as systems 1--9, as indicated in Figure~\ref{fig:civ} and Table~\ref{tab:lines} below.  Other \civ\ systems are present at -16,800 and -33,800~\kms\ in the spectrum, but they have narrow widths, complete covering, and blending problems in the \lyalpha forest, which, along with their high velocities, make them likely candidates for intervening gas and exclude them from further analysis in this work.  

After identifying the \civ\ doublets, we search the spectrum for other common NALs such as \siiv, \nv, \ciii, \ovi, and \hi\ Lyman series lines at the same redshift.  We also search for lower ionisation species, such as C~\textsc{II} and Si~\textsc{II}, but find none.  All of the systems, except possibly system 1, appear to have relatively high ionisations based on the presence and absence of high and low ionisation species respectively.  Each set of absorption lines at one redshift is considered a system, as labelled in Figure~\ref{fig:civ}.  Several of these systems are blends of two or three components, which are not individually labelled in the figure.

Systems 1 and 2 (Figures~\ref{fig:1} and~\ref{fig:2}) appear to be
line-locked in \civ.  The velocity offset between the \lam1548 line in system 1 and the \lam1551 line in system 2 is remarkably small ($< 2$~\kms) compared to the FWHMs of these lines ($\sim 30$~\kms) and the velocity shifts from the quasar systemic, $\sim -1440$ and $\sim -1940$~\kms. If this overlap between the \civ\ lines in systems 1 and 2 represents a physical line-lock, where the velocities of the two systems are actually separated by exactly their doublet separation, and not a chance alignment in the spectrum (see \citet{Ganguly03} and \citet{Braun89} for full discussions of the possible physical nature of line-locking), then it provides evidence for these lines forming in a quasar outflow driven by radiation pressure (see \S~\ref{sec:loc} below).

\subsection{Line Fitting}
\label{subsec:fit}
We fit each NAL system with a Gaussian optical depth profile.  The narrowest absorption lines are at least 1.5 times broader than the spectral resolution and the other lines are significantly broader than this.  The absorption lines are, therefore, fully resolved, and such Gaussian optical depth profile fits are sufficient to determine accurate optical depths and covering fractions.  The optical depths and covering fractions are held constant across the width of each line profile.  Gaussian fits are actually essential to distinguish individual absorption features in the crowded \lyalpha forest, and also useful to disentangle blended absorption in other areas of the spectrum. Furthermore, Gaussian fits smooth over noise spikes and large optical depth and covering fraction uncertainties in the wings of the lines.  We also use Gaussian fits to simultaneously fit and lock together various parameters including redshift, doppler $b$~parameter, covering fraction and a 2:1 optical depth ratio based on oscillator strength ratios for doublets such as the \civ, \siiv, \nv, and \ovi.  

To measure accurate optical depths, we must consider the possible effects of partial coverage of the emission source by the absorbing gas.  The line of sight covering fraction affects the observed line intensity as follows: 

\begin{equation}
\centering
I_{\mbox{v}}=(1-C_{f})I_0+C_{f}I_0e^{-\tau_{\mbox{v}}}
\label{eq:cf}
\end{equation} where $0 \le C_{f} \le 1$ is the velocity dependant line of sight covering fraction, $I_0$ is the emitted (unabsorbed) intensity  and $I_{\mbox{v}}$ and $\tau_{\mbox{v}}$ are the observed intensity and line optical depth at each velocity shift v.  This equation assumes that the background light source has a uniform brightness given by I$_0$ and the foreground absorber is homogeneous with a single value of $\tau_{\mbox{v}}$.  The viability of this assumption is discussed by \citet{Hamann04} and \citet{Arav05}. We assume that all lines in a given multiplet have the same C$_{f}$ at a given velocity.  We do not explicitly attempt to distinguish between partial covering of the continuum source and of the BEL region as discussed by \citet{Ganguly99}.  However, we estimate from the SDSS spectrum that the \civ\ BEL peaks 20\% above the continuum, which implies that the BEL can only account for partial covering of 0.8 or higher.

We attempt to fit each system with the smallest possible number of Gaussian components.  This minimizes the number of free parameters and provides a more robust characterization of column densities, ionisations and abundances in absorbing regions whose internal velocities might be more complex than simple Gaussians (e.g., in outflows).  We fit each absorption line with a single Gaussian unless 1) the system clearly has multiple components distinguished by inflection points that stand out significantly above the noise fluctuations in the spectrum (e.g. system 7), or 2) a single Gaussian would miss a significant fraction, $\ga 25$\%, of the absorption line strength (e.g. system 4).   The $\sim 25$\% threshold is somewhat arbitrary, but it ensures that we achieve a good fit to the observed line and that significant portions of absorption (i.e., large enough to change the column density measurements) are not missed.  For these exceptional cases we use the minimum number of Gaussians possible to achieve an accurate fit to the data.  If a system is fit with two or more Gaussians, each Gaussian is labelled as a component.  We assume that the covering fraction is the same for all components in a given system, such that the optical depths in Equation~\ref{eq:cf} simply add together in regions of component overlap (see Hamann et al. in preparation for further discussion).  This simplifying assumption is well justified by the excellent fits to all the systems, with the possible exception of system 6, which we discuss in more detail in \S~\ref{sec:sys}.  All ions with Gaussian fits are shown along with their Gaussian optical depth profiles in Figures~\ref{fig:1}-\ref{fig:9}.  Badly blended members of, e.g., the Lyman-series lines are not used to constrain ionisation or abundance in \S~\ref{sec:ion} and are not shown.  We check the Gaussian results using direct integration and point-by-point measurements of $\tau_{\mbox{v}}$ and $C_f$ across the line profiles for several systems with either non-Gaussian profiles (System 8) or $C_f < 1$ (Systems 5 and 6) as discussed later in this section.  

The free parameters in the Gaussian fits are $C_f$, central optical depth, $\tau$, the Doppler $b$ parameter, and redshift for each component in each system.  We fit the \civ\ absorption lines first, then base the fit for other absorption lines on the \civ\ fit parameters.  To ensure we are analysing the same gas in different ions, we fix the redshift for all absorption lines in a system to the \civ\ redshift.  We further exclude unwanted contributions from blends or complex multi-phase gas by fixing the $b$ parameter of all ions with ionisations lower than \civ\ to that of \civ.  Higher ionisation lines, such as \nv\ and \ovi, are allowed to be broader.  However, we cap the $b$-values of the \hi\ profiles at 140\% of \civ.  This cap is important for the abundance analysis below (\S~\ref{sec:ion}) because it ensures that the derived \hi\ column densities do not include gas with dramatically different kinematics than \civ.  We choose to cap the \hi\ $b$-values at 140\% of the \civ\ $b$-values instead of the much higher percentage expected for purely thermal broadening because the widths of the \civ\ lines exceed the thermal widths expected for a gas photoionised by either the quasar or the inter-galactic UV spectrum.  Therefore, we assume the $b$-values are dominated by non-thermal broadening effects.  On the other hand, setting the cap at 140\% instead of something smaller, such as 100\%, allows for some contribution of thermal broadening to $b$ in the narrower systems (which would affect \hi\ more than \civ).  Overall, our fits to the Lyman lines should lead to reasonable but generously large estimates of the amount of \hi\ gas that coexists with \civ, and therefore, to conservatively low estimates of the C/H abundance.

As stated above, the covering fraction is a free parameter in the Gaussian optical depth fits of each doublet.  In cases where the best fit profile has $C_f < 1$, we repeat the fit with $C_f = 1$ to test the robustness of the $C_f < 1$ result.  We then compare how well each of the two fits with different values of $C_f$ match the data.  In cases where the fits are comparable, we assume $C_f = 1$, otherwise, the best fit is chosen.  For example, we confirm that the $C_f < 1$ fit follows the data in systems 5 and 6, as shown in Figure~\ref{fig:5+6}, whereas the $C_f = 1$ fit does not match the observed doublet ratios in \civ\ and \nv.  We assume $C_f = 1$ for all singlet lines.  The covering fraction for the Lyman lines is fixed at the \civ\ doublet covering fraction.  This is necessary because the observed line ratios within the Lyman series are too severely affected by blending in the \lyalpha\ forest to yield their own independent measures of $C_f$.  

Table~\ref{tab:lines} lists fit parameters for all of the absorption lines that yield useful constraints for the ionisation and abundance analysis described in \S~\ref{sec:ion}.  Absorption lines that are badly blended are not used in subsequent analysis, and are not listed in the table.  Each system is listed separately by redshift and velocity shift, where negative velocities denote gas moving towards the observer.  Only the stronger member of the \ovi\ doublet is listed, as \ovi\ is never used in the abundance analysis because of either line saturation or strong blending in these lines in all systems.  However, the strength of \ovi\ is still useful as an indicator of the ionisation of the gas.  In systems where \nv\ is not present, we list upper limits for the stronger member of the \nv\ doublet for completeness.  Table~\ref{tab:lines} lists the central wavelength ($\lambda$) and doppler $b$ parameter values along with column densities and rest equivalent widths (\textsc{REW}) derived from the Gaussian optical depth profile fits.   Systems 4 and 7 each have two blended components.  The values of $\lambda, b$, and $\log N$ are listed separately for these components in Table~\ref{tab:lines}, but the \textsc{REW}s, listed only with the first component data, apply to the entire blend.  We measure upper limits on \hi\ column densities in all cases where all the Lyman series absorption lines are blended with intervening absorption lines in the \lyalpha forest.  The same is true for singlet ions with upper limits on the column densities.  

We estimate uncertainties for the column densities by placing the continuum at the top and bottom of the noise around the fitted continuum, corresponding to the reasonable maximum/minimum values ($\sim 3~\sigma$ uncertainties) for continuum placement.  We measure $3~\sigma$ uncertainties for the \hi\ column densities of 0.18~dex on average.  The covering fraction is $C_f = 1$ for all the systems, unless otherwise noted in the footnotes of Table~\ref{tab:lines}, e.g. systems 5 and 6.  We estimate the uncertainty in the covering fraction derived from the Gaussian profile fits both formally and informally.  The formal uncertainties are estimated by propagating the error spectrum through the calculation of $C_f$.  However, these uncertainties are much smaller than the informal uncertainties, which are dominated by uncertainty in continuum placement.  We estimate covering fraction uncertainties due to continuum placement uncertainties by first shifting the continuum near each red doublet member up and down by the $3 \sigma$ continuum uncertainty, and then fitting the doublets with this new continuum.  The actual uncertainties are probably smaller than the uncertainties we derive in this way, because a similar shift in the continuum around both doublet members (a more likely occurrence) produces smaller changes in $C_f$.  We find $C_f = 0.7 \pm 0.15$ for system 5 and $C_f=0.7 \pm 0.20$ for system 6. If we fix the covering fraction in systems 5 and 6 at $C_f = 1$ instead of at the measured values, the column densities in all ions decrease by an average of 0.25~dex.



\begin{table}
\centering
\caption{Individual absorption lines.}
\begin{tabular}{llccccclc}
\hline
\label{tab:lines}

& $\#$ & z$_{abs}$ & ID $\lambda_{rest}$ & $\lambda$ & \textsc{REW }& $b$  & $\log N$ \\
& & v (\kms) & &\AA &\AA & \kms & \cmN & \\

\hline
& 1 & 3.42865 & Ly $\gamma$ 973 & 4307.03 & 0.162 & 27.8 & 15.05 \\
& & -1442 & C III 977 & 4326.88 & 0.076 & 10.8 & $\le$13.37 \\
& & & Ly $\alpha$ 1216& 5383.78 & 0.445 & 27.8 & 15.05 \\
& & & N V 1239  & 5486.30 & $\le$0.002 & 10.8 & $\le$12.57 \\
& & & C IV 1548 & 6856.43 & 0.034 & 10.8 & 12.99 \\
& & & C IV 1551 & 6867.81 & 0.018 & 10.8 & 12.99 \\
\\
& 2 & 3.42133 & C III 977 & 4319.73 & 0.015 & 12.0 & $\le$12.39 \\
& & -1938 & Ly $\beta$ 1026 & 4535.05 & 0.011 & 14.0 & $\le$13.23 \\
& & & Ly $\alpha$ 1216 & 5374.88 & 0.069 & 14.0 & $\le$13.23 \\
& & & N V 1239 & 5477.24 & 0.034 & 20.5 & 13.24 \\
& & & N V 1243 & 5494.83 & 0.018 & 20.5 & 13.24 \\
& & & C IV 1548 & 6845.11  & 0.017 & 10.0 & 12.66 \\
& & & C IV 1551 & 6856.47 & 0.009 & 10.0 & 12.66 \\
\\
& 3 & 3.41864 & Ly $\epsilon$ 938 & 4143.82 & 0.011 & 13.9 & $\le$14.26 \\
& & -2120 & Ly $\gamma$ 973 & 4297.29 & 0.035 & 13.9 & $\le14.26$ \\
& & & Ly $\alpha$ 1216& 5371.61 & 0.186 & 13.9 & $\le$14.26 \\
& & & N V 1239   &  5473.90 & $\le$0.006 &  9.5  & $\le$12.45 \\
& & & Si IV 1394 & 6158.52 & 0.023 & 7.2 & 12.48 \\
& & & Si IV 1403 & 6198.30 & 0.013 & 7.2 & 12.48 \\
& & & C IV 1548 & 6840.94 & 0.096 & 9.5 & 13.62 \\
& & & C IV 1551 & 6852.29 & 0.062 & 9.5 & 13.62 \\
\\
& 4 & 3.41775 & Ly $\epsilon$ 938 & 4142.98 & 0.059 & 15.4 & $\le$14.86 \\
& & -2182 &  Ly $\gamma$ 973 & 4296.43 & 0.164 & 15.4 & $\le$14.86 \\
& & & Ly $\alpha$ 1216 & 5370.53 & 0.589 & 15.4 & $\le$14.86 \\
& & & N V 1239 & 5472.79 & $\le$0.055 & 11.0 & $\le$12.28 \\
& & & Si IV 1394 & 6157.28 & 0.081 & 9.1 & 12.76 \\
& & & Si IV 1403 & 6197.09 & 0.046 & 9.1 & 12.76 \\
& & & C IV 1548 & 6839.56 & 0.215 & 11.0 & 13.56 \\
& & & C IV 1551 & 6850.91 & 0.132 & 11.0 & 13.56 \\

& & 3.41747 & Ly $\epsilon$ 938 & 4142.72 & * & 46.8 & $\le$14.64 \\
& & -2200 &  Ly $\gamma$ 973 & 4296.15 & * & 46.8 & $\le$14.64 \\
& & & Ly $\alpha$ 1216 & 5370.19 & * & 46.8 & $\le$14.64 \\
& & & N V 1239 & 5472.44 & * & 33.4 & $\le$13.45 \\
& & & Si IV 1394 & 6156.89 & * & 14.0 & 12.75 \\
& & & Si IV 1403 & 6196.66 & * & 14.0 & 12.75 \\
& & & C IV 1548 & 6839.12 & * & 33.4 & 13.67 \\
& & & C IV 1551 & 6850.47 & * & 33.4 & 13.67 \\
\\
\hline
\end{tabular}
\end{table}

\begin{table}
\begin{tabular}{llccccclc}
\multicolumn{3}{l}{Table \ref{tab:lines}. Continued}\\
&5$^a$  & 3.40391 & C III 977 & 4302.71 & 0.059 & 19.7 & $\le$13.05 \\
& & -3121 & O VI 1032 & 4544.53 & 0.275 & 25.3 & $\le$14.99 \\
& & & Ly $\alpha$ 1216 & 5353.70 & 0.176 & 32.2 & 13.91  \\
& & & N V 1239 & 5455.65 & 0.098 & 13.0 & 14.30 \\
& & & N V 1243 & 5473.18 & 0.075 & 13.0 & 14.30 \\ 
& & & C IV 1548 & 6818.13 & 0.141 & 15.0 & 14.12 \\
& & & C IV 1551 & 6829.45 & 0.105 & 15.0 & 14.12 \\
\\
& 6$^b$& 3.40196 & Ly $\gamma$ 973 & 4281.07 & 0.055 & 31.7 & 14.41 \\
& & -3254 & Ly $\beta$ 1026 & 4515.18 & 0.137 & 31.7 & 14.41 \\
& & & O VI 1032 & 4542.51 & 0.739 & 63.5 & $\le$15.54 \\
& & & Ly $\alpha$ 1216 & 5351.33 & 0.382 & 31.7 & 14.41 \\
& & & N V 1239 & 5453.24 & 0.284 & 50.0 & 14.57  \\
& & & N V 1243 & 5470.76 & 0.188 & 50.0 & 14.57 \\
& & & C IV 1548 & 6815.11 & 0.220 & 45.0 & 14.02  \\
& & & C IV 1551 & 6826.42 & 0.128 & 45.0 & 14.02 \\
\\
& 7 & 3.39936 & Ly $\delta$ 950 & 4178.26 & 0.008 & 27.3 & $\le$13.19 \\
& & -3430 & C III 977 & 4298.26 & 0.102 & 19.8 & $\le$12.36 \\
& & & N III 990 & 4354.49 & 0.023 & 19.8 & $\le$13.04 \\
& & & Ly $\alpha$ 1216 & 5348.17 & 0.295 & 27.3 & $\le$13.19 \\
& & & N V 1239 & 5450.02 & 0.310 & 18.0 & 13.50 \\
& & & N V 1243 & 5467.53 & 0.180 & 18.0 & 13.50 \\
& & & C IV 1548 & 6811.09 & 0.187 & 19.9 & 13.09 \\
& & & C IV 1551 & 6822.40 & 0.101 & 19.9 & 13.09 \\

& & 3.39835 & Ly $\delta$ 950 & 4177.31 & * & 60.0 & $\le$13.74 \\
& & -3496 & C III 977 & 4297.28 & * & 43.7 & $\le$13.19 \\
& & & N III 990 & 4353.49 & * & 43.7 & $\le$13.04 \\
& & & Ly $\beta$ 1026 & 4511.48 & * & 60.0 & $\le$13.74 \\
& & & Ly $\alpha$ 1216 & 5346.95 & * & 60.0 & $\le$13.74 \\
& & & N V 1239 & 5448.77 & * & 52.4 & 14.22 \\
& & & N V 1243 & 5466.27 & * & 52.4 & 14.22 \\
& & & C IV 1548 & 6809.53 & * & 43.7 & 13.62 \\
& & & C IV 1551 & 6820.83 & * & 43.7 & 13.61 \\
\\
& 8 & 3.37983 & Ly $\beta$ 1026 & 4492.48 & 0.135 & 149.0 & $\le$14.29 \\
& & -4763 & O VI 1032 & 4519.68 & 1.705 & 149.8 & 15.86 \\
& & & Ly $\alpha$ 1216 & 5324.43 & 0.766 & 149.0 & $\le$14.29 \\
& & & N V 1239 & 5425.82 & 1.045 & 155.5 & 14.88 \\
& & & N V1243 & 5443.26 & 0.643 & 155.5 & 14.88 \\
& & & C IV 1548 & 6780.85 & 0.807 & 155.5 & 14.41 \\
\\
& 9 & 3.35922 & Ly $\gamma$ 973 & 4239.50 & 0.039 & 58.4 & $\le$14.02 \\
& & -6186 & C III 977 & 4258.94 & $\le$0.158 & 50.0 & $\le$13.35 \\
& & & Ly $\beta$ 1026 & 4471.34 & 0.112 & 58.4 & $\le$14.02 \\
& & & O VI 1032 & 4498.41 & 0.413 & 42.2 & 14.51 \\
& & & Ly $\alpha$ 1216 & 5299.37 & 0.513 & 58.4 & $\le$14.02 \\
& & & N V 1239 & 5400.29 & 0.212 & 40.0 & 13.84 \\
& & & N V 1243 & 5417.64 & 0.115 & 40.0 & 13.84 \\
& & & C IV 1548 & 6748.94 & 0.156 & 40.4 & 13.35 \\
& & & C IV 1551 & 6760.14 & 0.083 & 40.4 & 13.35 \\

\hline
\multicolumn{9}{l}{$^a$ System 5 has covering fractions 
$C_f(\hi) = 0.70$, $C_f(\nv) = 0.67$}\\
\multicolumn{9}{l} {and $C_f(\civ) = 0.70$. }\\
\multicolumn{9}{l}{$^b$ System 6 has covering fractions $C_f(\hi)~=~1.0$, $C_f(\nv)~=~0.67$}\\
\multicolumn{9}{l}{and $C_f(\civ)~=~0.72$. }\\
\end{tabular}
\end{table}


\vspace{2mm}

We perform a simple test to determine the reliability of the $C_f < 1$ result from the Gaussian profile fits for systems 5 and 6.  We predict the shape of the longer wavelength \civ\ and \nv\ doublet members based on the intensity of the shorter wavelength member, combined with the 2:1 $\tau$-ratio derived from the oscillator strengths of each line.  The predicted shorter wavelength member will only match the data if $C_f = 1$.  These predictions are shown in Figure~\ref{fig:tau}.  The observed data for both doublet members are plotted with solid curves and the predicted longer wavelength doublet member is plotted with a dot-dashed curve.  The predicted shape of the longer wavelength member of \civ\ and especially of \nv\ is much weaker than the observed shape for system 5.  In system 6, the line centres of \civ, and more clearly \nv, are stronger in the observed data for the longer wavelength members than in the predictions.   We conclude that $C_f < 1$ for at least some portion of each of these lines in systems 5 and 6.  

The Gaussian fitting technique described above assumes a single covering fraction across the entire line.  However, we demonstrate with the above $\tau$-ratio analysis that covering fractions are not always constant across a single line profile.  To better account for this, and to determine if the guassian fits find reasonable average values for $C_f$, particularly for lines with very non-Gaussian shapes, we use a point-by-point method in addition to the Gaussian fitting method to determine $\tau_{\mbox{v}}$ and $C_f$ across the line profiles in several systems.  We fit systems 5 and 6, the two systems with $C_f < 1$, along with system 8, which has a very non-Gaussian profile.  We step across the absorption line, calculating average intensity in each small (a few times the resolution) regularly spaced sections of the spectrum, using the ratio of the intensities in the doublet in Equation~\ref{eq:cf} to measure $C_f$ and $\tau_{\mbox{v}}$ at each step.  The point-by-point fits are shown in Figures~\ref{fig:cf5} and \ref{fig:cf8}.  The solid curve shows the shorter wavelength doublet member, while the dot-dashed curve shows the longer wavelength doublet member.  The covering fraction at each point is represented by the filled circles.  The steps used for system 6 are three resolution elements wide, which is wide enough to smooth over the noise but narrow enough to avoid blending the wings and core of the line.   However, system 5 is narrow enough that using bins three resolution elements wide, or wider, across the wings of the line would blend too much information from the core and the continuum.  Furthermore, the spectrograph resolution could be blending the covering fraction in the wings of the line with the continuum.  Thus, for the narrow system 5, we measure only the 3 resolution element bin at the line core.  The step size for system 8 is four resolution elements.  This larger step size further smooths over noise, and can be used because the line is much broader than the other systems, lessening the impact of blending of the core and wings of the line.  We derive formal covering fraction uncertainties ($\sigma_{C_f}$), represented as error bars at each point in Figures~\ref{fig:cf5} and \ref{fig:cf8}.  The average \civ\ and \nv\ central covering fraction from the point-by-point method matches the \civ\ and \nv\ covering fraction derived from the Gaussian fitting method to within 10\% in systems 5 and 6.  

Based on this result, the results of the $\tau$-derived doublet ratio analysis, and uncertainties derived from the Gaussian fitting method, we are confident of the accuracy of the $C_f < 1$ measurements for systems 5 and 6, although the exact value of the covering fraction remains unknown beyond the fact that it is below 1.  The \civ\ and \nv\ column densities found in system 6 using the point-by-point method match the \civ\ and \nv\ column densities derived from the Gaussian fitting method to within 0.14~dex and 0.06~dex respectively. The same comparison for \nv\ in system 8 yields a difference of 0.18~dex between the two methods.  We conclude that the Gaussian technique is sufficient for comparing different systems and generally provides accurate column density and covering fraction results for the purposes of this work.

\subsection{Ionisation and Abundances}
\label{sec:ion}

The abundance ratios can be derived from the ratio of measured column densities corrected for the degree of ionisation in the gas.
For example, the relative carbon to hydrogen abundance normalised to
solar is given by:

\begin{equation}
\centering
\left[\frac{C}{H}\right]=\log\left(\frac{N(\civ)}{N(H I)}\right) + \log\left(\frac{f(H I)}{f(\civ)}\right) + \left[\frac{H}{C}\right]_{\sun}
\label{eq:c/h}
\end{equation} where $f$ is the ionisation fraction of a given ion, $N$ is the column
density and the final term on the right-hand side is the logarithmic
solar abundance ratio of hydrogen to carbon listed in \citet{Grevesse07}. The second term on the right is the ionisation correction (IC).  These correction factors can be large when comparing a highly ionised metal like \civ\ to \hi.  The exact values depend on the ionisation mechanism.  Photoionisation by the quasar spectrum is by far the most likely scenario based on the arguments in \S~\ref{sec:loc} that all of the systems are likely to be intrinsic to the quasar environment.

We derive values of IC using the photoionisation calculations shown in \citet{Hamann10}.  Their calculations adopt a nominal quasar spectrum consistent with recent observational estimates at the critical ionising (far-UV) photon energies\footnote{The calculations in \citet{Hamann10} apply to gas that is photoionised by
a typical quasar spectrum. We perform additional CLOUDY \citep{Ferland98} calculations using the inter-galactic background spectrum in CLOUDY, which is based on Haardt \& Madau (2005, private communication). We find that the ionisation fractions of interest in the present work have only negligible differences between the two calculations, e.g., compared to uncertainties in the measured quantities or derived ionisation constraints. Therefore, our analysis of the ionisation and
abundances in J1023+5142 should apply whether the absorbers are located near the quasar or outside the quasar's radiative sphere of influence.}. The calculations also assume that the absorbing gas is optically thin in the Lyman continuum, which is appropriate for the column densities we measure in the absorption lines of J1023+5142 (Table~\ref{tab:lines}).

Ideally, we would constrain the absorber ionisations by comparing the ratios of observed column densities in different ions of the same element, such as $N(\ciii)/N(\civ)$ or $N(\niii)/N(\nv)$, to the theoretical results in \citet{Hamann10}.  However, these constraints are only marginally usable in our data because  $N(\ciii)$ and $N(\niii)$ are always blended in the \lyalpha forest and are therefore only ever constrained as upper limits.  Therefore we estimate the IC from ratios such as $N(\nv)/N(\civ)$ or $N(\siiv)/N(\civ)$, with the additional assumption that the relative metal abundances are approximately solar.  The specific ionisation constraints used for each system sometimes lead to upper limits, lower limits or specific values for the abundance ratios, and are described in more detail for individual systems in \S~\ref{sec:sys} below.  Our best estimates for the $C/H$ abundances based on these constraints are all super--solar, except in system 1.  Table~\ref{tab:abs}, which contains several different abundance indicators for each NAL system, lists these estimated 'best' abundances in column 3, titled $[C/H]_{best}$, for the nine systems.  
   
We also calculate robust lower limits on the metal to hydrogen abundance ratios by applying minimum values of the ionisation correction (IC$_{min}$, \citet{Hamann97}) to the measured \civ, \siiv\ and \nv\ column densities, when available.  Each metal ion has a unique global IC$_{min}$ that occurs near the peak of its own ionisation fraction.  For example, $f(\hi)/f(\civ)$ peaks approximately where $f(\civ)$ is largest.  We use the values of IC$_{min}$ listed in \citet{Hamann10}.   Applying these minimum correction factors to the observed column density ratios (Equation~\ref{eq:c/h}) leads to the firm lower limits listed for $[C/H]_{min}, [Si/H]_{min}$ and $[N/H]_{min}$ abundances in columns 4--6 of Table~\ref{tab:abs}.  The minimum ionisation corrections provide firm lower limits on the abundances that do not depend on the ionisation uncertainties or the possibility of a multi--phase gas.  In particular, any gas components not at an ionisation corresponding to IC$_{min}$ would have the effect of raising the actual value of IC and thus also the actual abundance. 

We derive total H column densities for each NAL system from the \hi\ column densities listed in column 7 of Table~\ref{tab:lines} and the best ionisation correction described above.  We use \begin{equation} \log N(H) = \log N(\hi) - \log f(\hi),\end{equation} where $\log N(H)$ is the total H column density, $\log N(\hi)$ is the column density of \hi\ and $\log f(\hi)$ is the \hi\ fraction used to obtain IC.  $\log N(H)$ for each system is listed in column 7 of Table~\ref{tab:abs}.  

The uncertainties in these results are dominated by uncertainties in the IC.  In addition to the limited constraints provided by the data, a few well-studied cases have shown that individual absorbers can span a range of ionisations and have a range of IC values (e.g. \citet{Hamann97}).  We assume a single ionisation state for each absorption line system.   We discuss the individual systems briefly in \S~\ref{sec:sys}.  
 
\begin{small}
\begin{table}

\centering
\begin{minipage}{240mm}
\caption{Metal Abundance and Total H Column Density}
\begin{tabular}{lccccccc}
\hline
\label{tab:abs}
 & \# & z$_{abs}$ & [C/H]$_{best}$ & [C/H]$_{min}$ & [N/H]$_{min}$ & [Si/H]$_{min}$  & log(N(H))~cm$^{-2}$\\
\hline
&  1 & 3.42865 & $\le$-0.47 & $\ge$-2.25 & -- & --  & 17.62 \\
& 2 & 3.42133 & $\ge$+0.32 & $\ge$-0.76 & $\ge$-0.47 & --  & $\le$17.95\\
& 3 & 3.41864 & $\ge$+0.38 & $\ge$-0.82 & -- & $\ge$-0.34  & $\le$17.16 \\
& 4 & 3.41775 & $\ge$+0.79 & $\ge$-1.32 & -- & $\ge$-0.66  & $\le$17.24\\
& 5 & 3.40391 & +0.72 & $\ge$+0.02 & $\ge$-0.10 & --  & 18.21 \\
& 6 & 3.40196 & +0.50 & $\ge$-0.58 & $\ge$-0.32 & --  & 19.13 \\
& 7 & 3.39889  & $\ge$+0.40 & $\ge$-0.30  & $\ge$+0.01 & -- & $\le$18.19 \\
& 8 & 3.37985 & $\ge$+0.94 & $\ge$-0.07 & $\ge$-0.17 & --  & $\le$18.90 \\
& 9 & 3.35906 & $\ge$+0.14 & $\ge$-0.86 &$\ge$ +0.11 & --  & $\le$18.60 \\
\hline
\end{tabular}
\end{minipage}

\end{table}
\end{small}

\section{Notes on Individual Systems}
\label{sec:sys}

{\it \noindent System 1, $\mbox{v} = -1441$~\kms}

\begin{figure}
\centering
\vspace{4mm}
     \includegraphics[width=7cm]{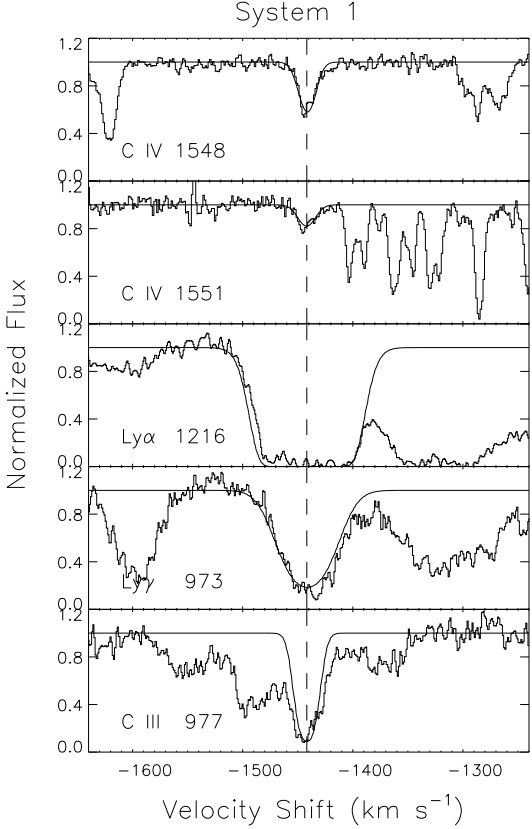}
     \caption{Line profiles in the normalised spectrum J1023+5142 for system 1.  The central velocity for the Gaussian profile fit is $\mbox{v} = -1441$~\kms. The velocity scale is with respect to the rest frame of the quasar based on $z_{em} = 3.45$, where negative velocities denote motion towards the observer and away from the quasar. The velocity range is 400~\kms\ for this and figures \ref{fig:2} through \ref{fig:7} and \ref{fig:9}.  The solid curve in each panel is the Gaussian optical depth fit to individual lines.  The dashed vertical line is the central velocity of the system.   All of the lines used to derive or constrain column densities with Gaussian fits are shown in the figure. }

\label{fig:1}
\end{figure}

The \civ\ in system~1 appears line-locked with the \civ\ in system~2, as discussed further in \S~\ref{sec:loc}.  \ovi\ is not present, or is very weak, implying that the ionisation is low.  Further evidence for low ionisation is the weak \civ\ combined with strong \hi\ measured in \lyalpha and Lyman~$\gamma$, as seen in Figure~\ref{fig:1}. Our best ionisation constraint comes from an upper limit on \ciii, which means the best $C/H$ abundance is an upper limit as well.  This gas is a likely candidate for host galaxy halo gas based on the weakness of the metal lines and the low abundances.

\vspace{0.1in}
{\it \noindent System 2, $\mbox{v} = -1938$~\kms}

\begin{figure}
\centering
\vspace{4mm}
 \includegraphics[width=7cm]{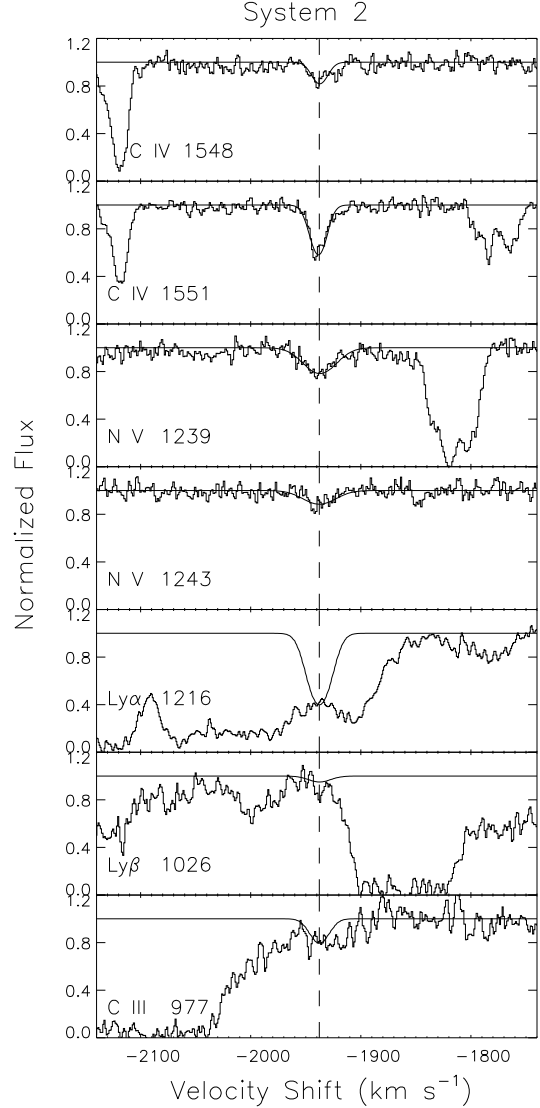}
\caption{Line profiles in the normalised spectrum J1023+5142 for system 2.  The central velocity for the Gaussian profile fit is $\mbox{v} = -1938$~\kms.  The symbols and ranges are the same as in Figure~\ref{fig:1}.}
\label{fig:2}
\end{figure}

The \civ\ in system 2 appears to be line-locked with
system~1, as mentioned above and discussed further in \S~\ref{sec:loc}.  The \hi\ could be shifted to a lower velocity by as much as 30~\kms\ from the metal lines in this system, indicating a multi--phase gas, but heavy blending obscures the precise shift of the lines as can be seen in Figure~\ref{fig:2}.  The \lyalpha\ absorption line is poorly constrained.  The resulting \hi\ optical depth and doppler $b$ parameter are upper limits, resulting in lower limits for the best estimate of $C/H$ abundance.  We constrain the ionisation by the relative strengths of \civ\ and \nv, assuming solar abundance
ratios. 

\vspace{0.1in}
{\it \noindent System 3, $\mbox{v} = -2120$~\kms}

\begin{figure}
\vspace{4mm}
\centering
   \includegraphics[width=7cm]{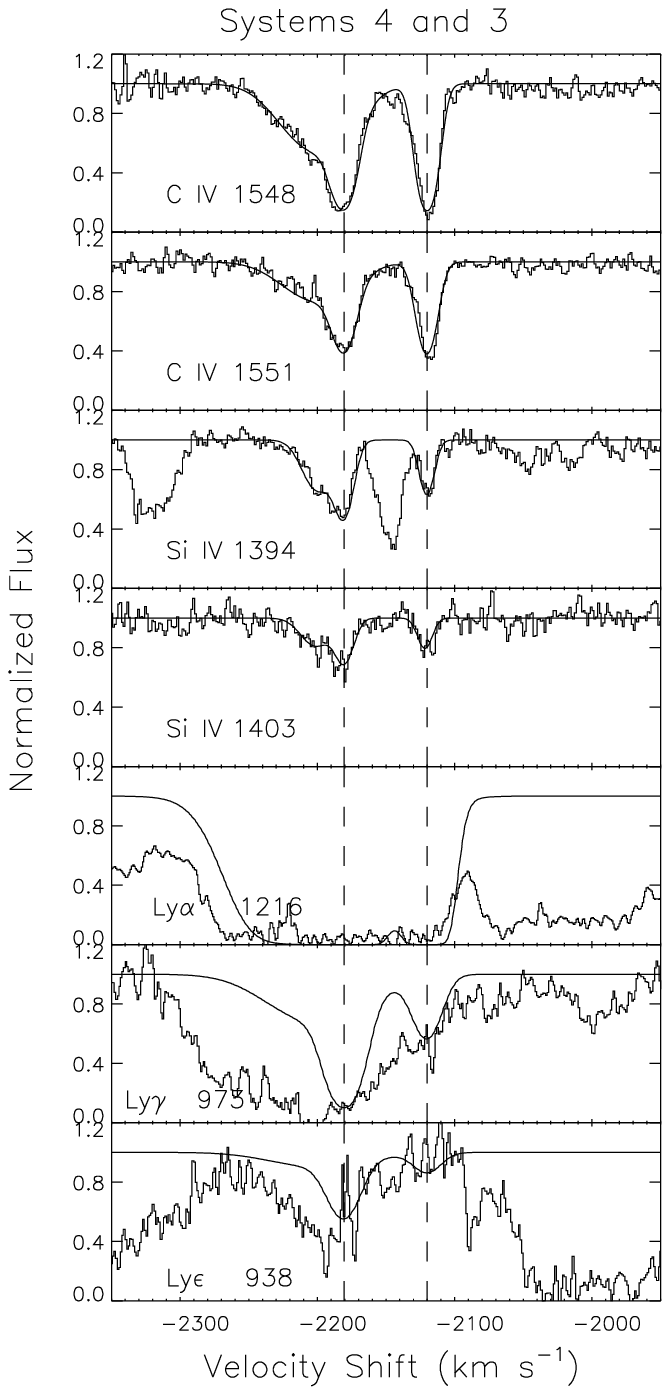}
\caption{Line profiles in the normalised spectrum J1023+5142 for systems 3 and 4.  The central velocities of the Gaussian profile fits are $\mbox{v} = -2120$~\kms\ for system 3 and $\mbox{v} = -2200$~\kms\ for system 4.  The symbols and ranges are the same as in Figure~\ref{fig:1}. }
\label{fig:3+4}
\end{figure}
The \hi\ lines in system 3 are blended with those from system 4, but appear consistent with the metal lines, shown in
Figure~\ref{fig:3+4}.  Because of the relatively poor constraints on the \hi\ absorption lines, the \hi\ optical depth and doppler $b$ parameter are upper limits, resulting in lower limits for the best estimate of $C/H$ abundance.  We constrain the ionisation by the relative strengths of \civ\ and \siiv, assuming solar abundance ratios.  

\vspace{0.1in}
{\it \noindent System 4, $\mbox{v} = -2182, -2200$~\kms}

The \civ\ and \siiv\ doublets in system 4 are fit with two blended
Gaussian components to accommodate the asymmetric profile.  We use the central velocity of each component to identify the system.  The \hi\ absorption lines are poorly constrained due to blending with system 3.  The resulting \hi\ optical depth and doppler $b$ parameter are upper limits, resulting in lower limits for the best estimate of $C/H$ abundance.  We constrain the ionisation by the relative strengths of \civ\ and \siiv, assuming solar abundance ratios.  This system is broad and asymmetric, which is indicative of a wind or outflow feature (see \S~\ref{sec:loc}).

\vspace{0.1in}
{\it \noindent System 5, $\mbox{v} = -3121$~\kms}

\begin{figure}
\centering
\vspace{4mm}
     \includegraphics[width=7cm]{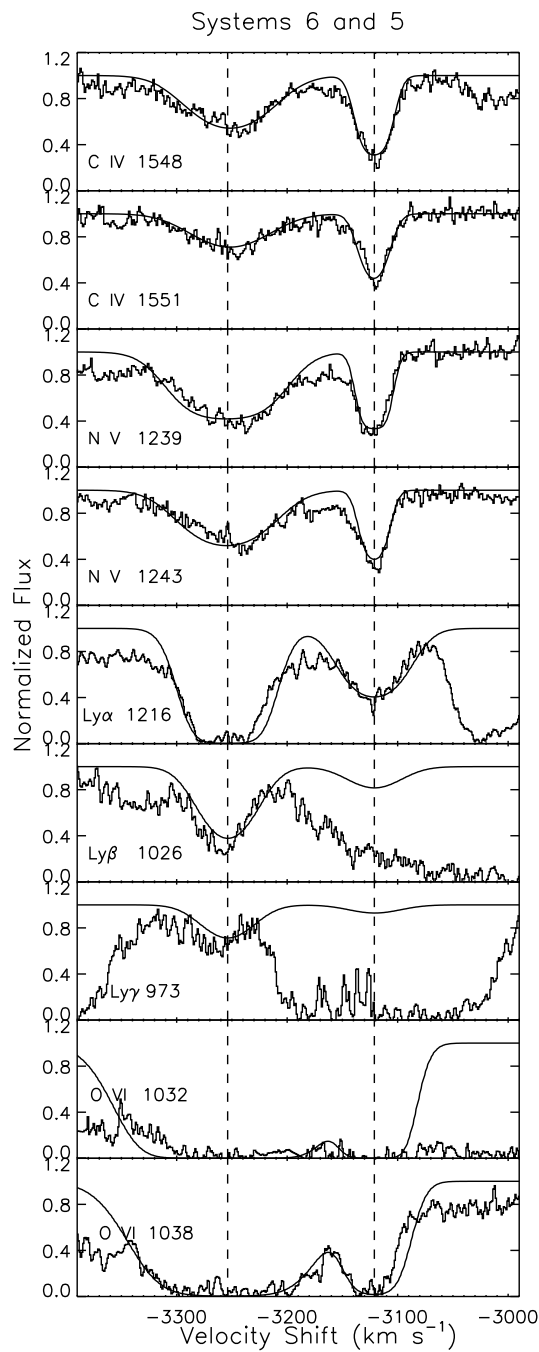}
     \caption{Line profiles in the normalised spectrum J1023+5142 for systems 5 and 6.  The central velocities of the Gaussian profile fits are $\mbox{v} = -3121$~\kms\ for system 5 and $\mbox{v} = -3254$~\kms\ for system 6.  The solid curve is the $C_f < 1$ Gaussian fit for the \civ, \nv\ and \hi\ absorption lines.  System 5 is the narrower system.  The ranges are the same as in Figure~\ref{fig:1}. }
\label{fig:5+6}
\end{figure}
\begin{figure}
\centering
\vspace{4mm}
\includegraphics[width=7cm]{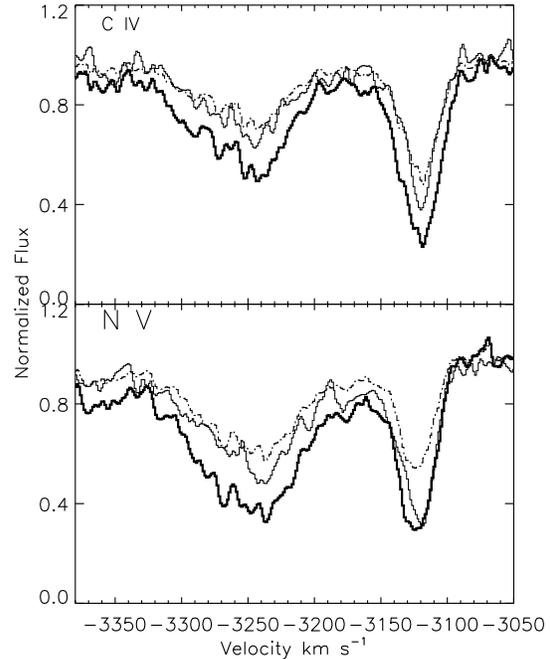}
\caption{$\tau$-predicted line profiles for systems 5 and 6. System 5 is the narrower system.  The dot-dashed curve shows the smoothed \civ\ and \nv\ predicted long wavelength doublet member, based on doublet optical depth ratio from short wavelength member, assuming $C_f = 1$.  The actual smoothed data for the shorter and longer wavelength doublets are shown as the bold and thin solid curves, respectively.  The longer wavelength data are stronger than the predictions, indicating partial covering, especially for the \nv\ doublet in system 5 and in the centre of the \nv\ doublet in system 6. }
\label{fig:tau}
\end{figure}
\begin{figure}
\centering
\vspace{4mm}
\includegraphics[width=7cm]{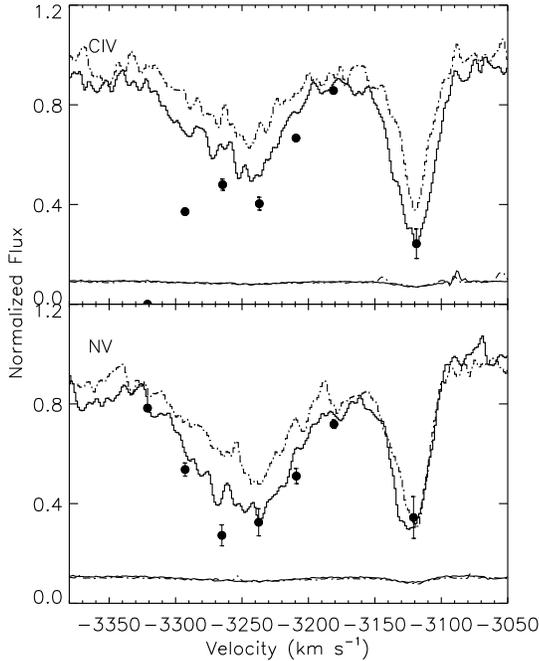}
\caption{Point-by-point covering fractions for \civ\ and \nv\ in system 6 and the centre of system 5 with step size of three resolution elements.  System 5 is the narrower system.  The solid curve is the smoothed shorter wavelength line, the dashed curve is the smoothed longer wavelength line, with their respective error spectra below.  The circles represent $1 - C_f$ at each step so that a point at zero flux has complete coverage, and a point at the continuum flux of one has no coverage. The circles are located at the centre of the average velocity steps.  }
\label{fig:cf5}
\end{figure}

The \hi\ in system 5 is well constrained by \lyalpha.  The \ovi\ is
strongly blended with that of system 6 as shown in
Figure~\ref{fig:5+6}.  The covering fraction in the doublet is
$\sim 0.7$.  The $C_f = 0.7$ Gaussian fits are shown as solid curves in Figure~\ref{fig:5+6}.  System 5 appears to have 2 components; a narrow, optically thick component sitting directly on top of a broader, optically thin one.  This is most clearly seen in
Figure~\ref{fig:5+6} in the longer wavelength members of the \civ\ and
\nv\ doublets, which have a much sharper central feature than their
shorter wavelength counterparts.  We constrain the ionisation by the
relative strengths of \civ\ and \nv, assuming solar abundance ratios.
The partial coverage in this system indicates that it is intrinsic to the quasar.  The partial coverage in this system and in system 6 are examined qualitatively with the $\tau$-ratio predicted doublets, shown in Figure~\ref{fig:tau}, and further with the point-by-point analysis, illustrated in Figure~\ref{fig:cf5}.  Both analyses confirm similar $C_f < 1$ results in both systems (See \S~\ref{subsec:fit} for details).

\vspace{0.1in}
{\it \noindent System 6, $\mbox{v} = -3254$~\kms}

\hi\ is well-constrained in system 6, with three mostly blend-free Lyman lines. The covering fraction in the doublets is $C_f \sim 0.7$, similar to system~5. The solid curves in Figure~\ref{fig:5+6} represent the $C_f < 1$ Gaussian fits, as for system 5.  The \ovi\ lines are blended with the \ovi\ lines in system~5.  The covering fraction in \hi\ appears to be $C_f = 1$ because \lyalpha\ reaches zero intensity. We constrain the ionisation by the relative strengths of \civ\ and \nv, assuming solar abundance ratios. The broad smooth shape, along with the partial coverage indicate that this system is part of an outflow.

This system appears somewhat asymmetric and the $\tau$-ratio analysis in Figure~\ref{fig:tau} suggests further that there may be two components, one with partial covering near the line-centre, and a second broader component with complete covering in the blue wing.  Although one component does not provide the best possible fit to all the lines in system 6, it is not clear that adding a second distinct component would provide a better characterisation of the actual conditions in the absorber.  We test this by fitting the system with one and two Gaussian components, where the two component fit still assumes the same covering fraction in both components.  Both fits produce similar column densities in all ions, $\Delta N(\civ) = 0.15$~dex, $\Delta N(\nv) = 0.1$~dex and $\Delta N(\hi) = 0.1$~dex in the same direction, therefore we prefer the single Gaussian fit in keeping with our prescription to minimise free parameters in the fits.  Also, by using the single Gaussian fit, we ignore parts of the \lyalpha\ absorption which do not correspond directly to \civ\ absorbing gas, and therefore retain the ability to directly compare \hi\ and \civ\ column densities for the abundance analysis.  

\vspace{0.1in}
{\it \noindent System 7, $\mbox{v} = -3430, 3496$~\kms}

\begin{figure}
\vspace{4mm}
\centering
     \includegraphics[width=7cm]{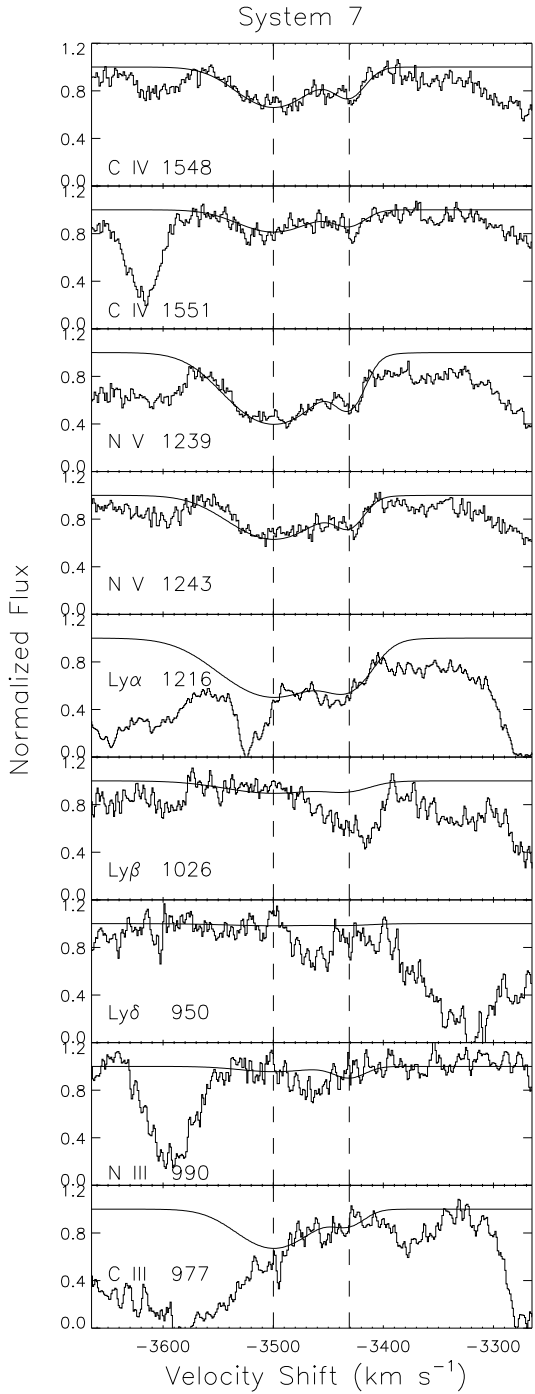}
\caption{Line profiles in the normalised spectrum J1023+5142 for system 7.  This system has two blended components with central velocities of $\mbox{v} = -3430$ and -3496~\kms.  The symbols and ranges are the same as in Figure~\ref{fig:1}. }
\label{fig:7}
\end{figure}

The \hi\ column density is constrained as an upper limit in system 7 because of blending in the Lyman lines, shown in Figure~\ref{fig:7}, resulting in lower limits for the  best estimate of $C/H$ abundance.  We fit this broad system with two Gaussian components to better match the absorber shapes, and identify the system by the central velocities of the two components. The ionisation is constrained by the relative strengths of \civ\ and \nv, assuming solar abundance ratios. 

\vspace{0.1in}
{\it \noindent System 8, $\mbox{v} = -4763$~\kms}

\begin{figure}
\centering
\vspace{4mm}
 \includegraphics[width=7cm]{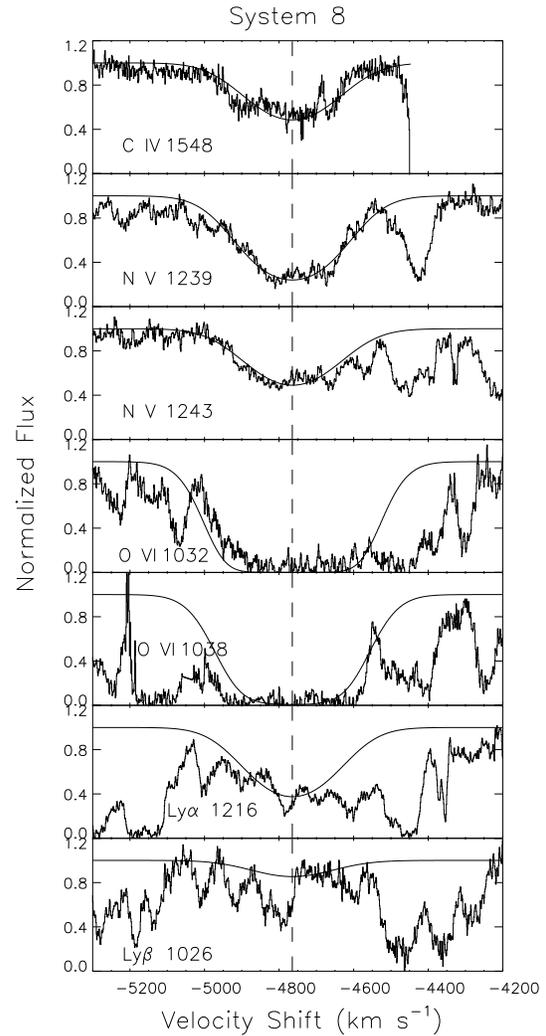}
\caption{Line profiles in the normalised spectrum J1023+5142 for system 8.  The Gaussian profile fit has a central velocity of $\mbox{v} = -4763$~\kms.  The velocity range is 1100~\kms.  The symbols and ranges are the same as in Figure~\ref{fig:1}.}
\label{fig:8}
\end{figure}

The longer wavelength member of the \civ\ doublet in system 8 falls on a gap between orders of the spectrograph between 6785 and 6795~\AA, but the \nv\ doublet is present in the spectrum, as is the shorter wavelength member of the \civ\ doublet.  The \nv\ doublet is used to determine the $C_f$ and the Doppler $b$ parameter for both doublets.  This system is almost broad enough to be a mini-BAL, and is likely an outflow system based on the shape and strength of the line profile, shown in Figure~\ref{fig:8}.  The \hi\ appears to be relatively weak in this system compared to the metal lines, although there is severe blending in the \lyalpha\ forest.  This blending means the \hi\ absorption is poorly constrained with an upper limit, and therefore the best estimate for $C/H$ abundance is a lower limit.  We constrain the ionisation by the relative strengths of \civ\ and \nv, assuming solar abundance ratios.

\begin{figure}
\vspace{4mm}
\centering
     \includegraphics[width=7cm]{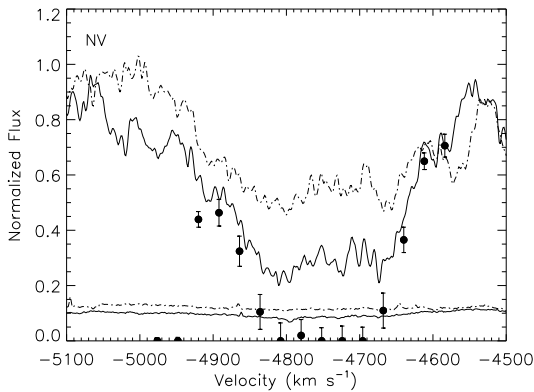}
     \caption{Point-by-point covering fractions for \nv\ in system 8 with step size of four resolution elements.  The symbols are the same as in Figure~\ref{fig:cf5}.}
\label{fig:cf8}
\end{figure}

We use the Gaussian fit to compare system 8 to other systems, but
the profile of system 8 is distinctly non-Gaussian.  Therefore we also fit the central trough of the line with a point-by-point analysis, shown in Figure~\ref{fig:cf8}.  The $C/H$ abundance found by the Gaussian fit is consistent within 10\% of the $C/H$ abundance found using the point-by-point method.

\vspace{0.1in}
{\it \noindent System 9, $\mbox{v}=-6083, -6186, -6298$~\kms}

\begin{figure}
\vspace{4mm}
\centering
     \includegraphics[width=7cm]{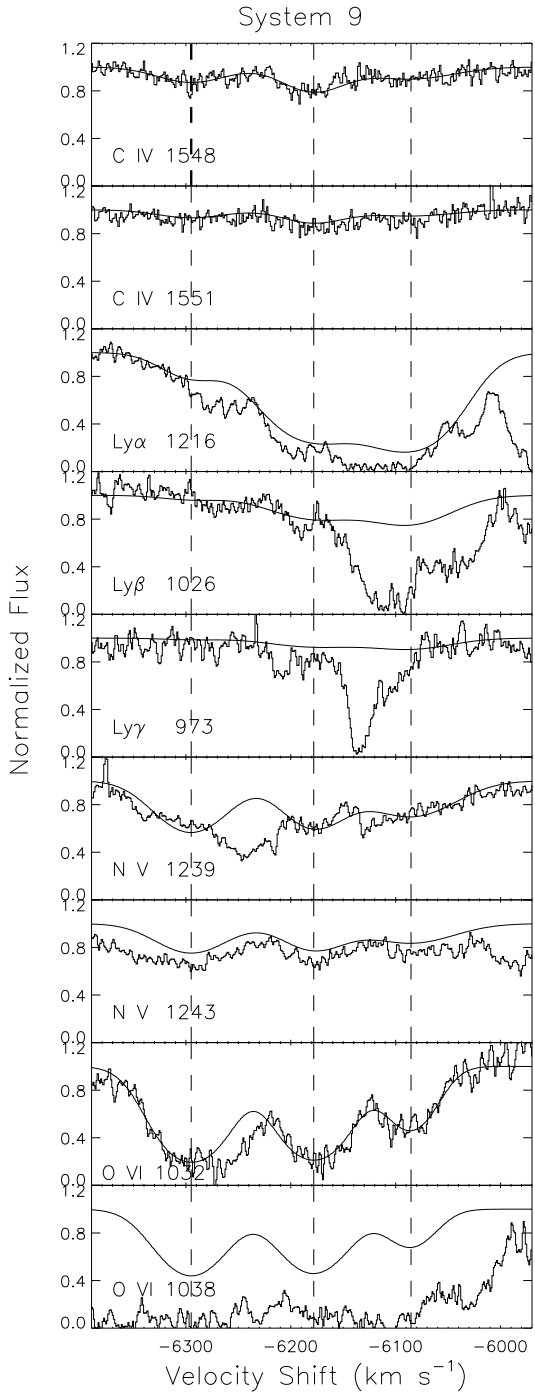}
     \caption{Line profiles in the normalised spectrum J1023+5142 for system 9.  This system is a blend of three components with central velocities, $\mbox{v} = -6083$, -6186 and -6298~\kms.   Although all three components are fit with Gaussian profiles, only the central component is considered in the abundance analysis.  The symbols and ranges are the same as in Figure~\ref{fig:1}.}
\label{fig:9}
\end{figure}
System 9 has three components, but we chose to analyse only the central component for abundances, as the two outer components are very poorly constrained, as shown in Figure~\ref{fig:9}.  This system has the highest velocity shift out of the group of narrow absorption lines, and lies just nominally outside of the velocity shift region for associated lines (v $\geq -5000$~\kms), at $\sim -6200$~\kms.  The Lyman lines could be shifted up to 20~\kms\ from the metal lines, indicating a possible multiphase gas, but the line are too weak to determine their precise centroids.  The weakness of the Lyman lines, along with blending in the \lyalpha\ forest mean the \hi\ column densities are upper limits, so the best estimate of the $C/H$ abundance is a lower limit.  We constrain the ionisation by the relative strengths of \civ\ and \nv, assuming solar abundance ratios.

\section{Discussion}
\label{sec:disc}

J1023+5142 has nine NAL systems with a range of column densities from $N(H) \leq 10^{17.2}$ to $10^{19.1}$~cm$^{-2}$, velocities from --1400 to --6200~\kms, \civ\ doppler $b$ values from 7 to 150~\kms, \civ\ \textsc{REW}(1548\AA) from 0.02 to 0.81~\AA\ and two systems with partial covering of either the continuum source or the broad emission line region (BLR), $C_f \approx 0.7$, which imply absorber diameters of $\leq 0.03$~pc or $\leq 0.8$~pc (discussed below in \S~\ref{sec:out}).  These systems are generally much weaker than those studied in larger statistical surveys of NALs, such as \citet{Vestergaard03}, which use lower resolution data, and measure \civ\ \textsc{REW} integrated across the doublet, with completeness limits of 0.3--0.5~\AA.  The NAL systems all appear to be highly ionised; none of the systems exhibit low ionisation species such as Si~\textsc{II}, C~\textsc{II} or Si~\textsc{III}, whereas all contain \civ\ and some contain higher ionisation species such as \ovi\ and \nv.  Systems 5, 6, 8 and 9 exhibit high ionisation (\ovi) absorption, and others may also have absorption at these wavelengths that is not observable due to blending in the \lyalpha\ forest.  Systems 2--9 exhibit supersolar metallicities ranging from $Z \ge 1$ to $\ge 8~Z_{\odot}$.  System 1 has a slightly lower metallicity of $Z \leq 0.3~Z_{\odot}$. We examine several diagnostics to estimate directly the location of each system.

\subsection{Location of the Gas}
\label{sec:loc}

The tight grouping and similar high metallicities (see \S~\ref{sec:metal} for further discussion) for all but one (systems 2--9) of the nine \civ\ absorption line systems in J1023+5142 suggest a possible physical connection between the absorbers.  The proximity of this NAL complex to the quasar redshift suggests further that the physical relationship includes the quasar itself. The velocity span across the group is too large to be explained by a single galaxy or even a large cluster of galaxies. It might be consistent with some larger cosmic structure connected to the quasar, but then we would expect the velocity distribution to include the red side of the quasar systemic. A more likely explanation is that the NAL complex formed in a multi-component outflow from the quasar.

There are several indirect arguments for an intrinsic origin for the gas in this NAL complex.  i) 8 of the 9 systems have super--solar metallicities, discussed in detail in \S~\ref{sec:metal} below.  ii) Some authors have argued that strong \ovi\ absorption may indicate intrinsic gas near the quasar.  \citet{Fox08} carry out a detailed study of \ovi\ absorption in $2 <  z < 3$ quasars and argue that $\log N(\ovi) \geq 15.0$~log(\cmN) indicates an intrinsic origin, supported by evidence for partial covering in most of these systems.  We measure $N(\ovi)$ in four systems in J1023+5142.  Two of them (6 and 8) are above the intrinsic thereshold defined by Fox et al., while the other two (5 and 9) are very near this threshold at  $\log N(\ovi) \geq 14.5$.  iii) All systems with \ovi, that is systems 5, 6, 8 and 9, also have strong \nv\ compared to \civ.  Strong \nv, especially compared to \civ, is often (though not always) present in intrinsic gas (e.g. \citet{Weymann81, Hartquist82, Hamann97c, Kuraszkiewicz02, Fox08}).  iv) The presence of strong \ovi\ and \nv, especially with the absence of low ionisation species such as C~\textsc{II} in these absorption systems is consistent with gas exposed to the intense ionising radiation field near a quasar.

If the NALs in J1023+5142 are intrinsic to the quasar environment, the most likely origin is in a quasar--driven outflow.  Other possible intrinsic origins all have lower velocities:  i) starburst-driven outflows typically have $100 < \mbox{v} < 1000$~\kms\ \citep{Heckman00}, and in Seyfert galaxies have maximum outflow speeds of 600~\kms, and more typical speeds of 100-200~\kms\ \citep{Rupke05b}, ii) other galactic/halo gas should have velocities near the typical velocity dispersion for such galaxies ($\sigma \sim 300$~\kms), iii) gas in the narrow line region of the quasar has typical velocities of $\mbox{v} \leq 1000$~\kms, and maximum velocities of $\mbox{v} \leq 2000$~\kms~ \citep{Ruiz01, Ruiz05, Veilleux05}, and iv) intra-cluster galaxy motions are shown by \citet{Popesso06} to generally have velocity dispersions $\sigma_{\mbox{v}} < 1000$~\kms\ or less for clusters with higher numbers of active galactic nuclei \citep{Richards99, Heckman00, Vestergaard03, Nestor08}. 

Statistical studies, \citet{Nestor08} (see also \citet{Wild08}), have shown that $> 43$\% of NALs at $-750 \ge \mbox{v} \ge -12000$~\kms\ with \textsc{REW}(1548\AA) $> 0.3$~\AA\ form in high-velocity quasar outflows.  This percentage increases to $\sim$57\% for the narrower range of $-1250 \ge \mbox{v} \ge -6750$~\kms, spanned by the NALs in J1023+5142.  The percentage reaches $\sim$72\% for the narrow range of $-1250 \ge \mbox{v} \ge -3000$~\kms, which encompasses systems 1 through 4 in J1023+5142.  These percentages are probably lower for weaker lines (\citet{Nestor08} and private communication). \citet{Misawa07a} also find that for \civ\ NALs with \textsc{REW}(1548\AA) $> 0.056$~\AA\ at velocities $v <  5000$~\kms\ the intrinsic (outflow) fraction is $\ge 33$\% and at higher velocities, $5000 < v < 70000$~\kms, the intrinsic fraction is $\ge 10-17$\%.  Nonetheless, these outflow fractions support the idea that most or all of the systems in this group of nine NALs in J1023+5142 form in a quasar outflow.  

We search for direct signatures of quasar outflow origin via 1) line variability, 2) partial covering and 3) broad profiles (see \citet{Hamann10, Hamann08} and references therein for more discussion).  

1) We have only very poor constraints on the variability.  We compare \civ\ and \nv\ \textsc{REW} results measured from the guassian fits to the SDSS and Keck spectra ($\Delta t_{rest} \approx 11$~months) in search of variability in the absorption lines.  System 8 is the only individual \civ\ and \nv\ system resolved in the SDSS spectrum, while the weaker lines are not detected in the SDSS spectrum.  System 8 is the strongest of the nine systems, and did not vary in \textsc{REW} by more than 15\% in \civ\ and \nv\ between the SDSS and Keck observations.  For the eight weaker systems, we conclude only that variability greater than a factor of 2 to 3 did not occur. 

2) There is partial covering in two (systems 5 and 6, Figures~\ref{fig:5+6} and \ref{fig:cf5}) out of the nine systems.  Absorption lines with partial covering of the luminosity source are attributed to gas near the quasar because partial covering is not expected to occur in intervening clouds or galaxies \citep{Hamann10}. The presence of partial covering in these lines strongly suggests that the gas is intrinsic and located in the near quasar environment. 
 
3) The profiles of systems 4, 6, 7, 8, and 9 shown in Figures~\ref{fig:3+4}, \ref{fig:5+6}, \ref{fig:7}, \ref{fig:8} and \ref{fig:9} have \civ\ and \nv\ $b$ values between 33 and 155~\kms and \ovi\ $b$ values between 42 and 150~\kms.  These $b$ values are broad and smooth compared to the thermal widths for gas at the highest expected temperature near $T~=~10^5$~K \citep{Arnaud85, Hamann95} for a photoionised gas near a quasar (33~\kms\ for H and less than 10~\kms\ for C and N).  They are also broader than typical non-damped \lyalpha\ intervening \civ, \nv\ and \ovi\ absorption lines, which have on average $b < 12-14$~\kms\ for \ovi, and $b < 10-12$~\kms\ for \civ\ and \nv~ \citep{Tzanavaris03, Bergeron05, Schaye07, Fox08}.  These profiles, therefore, exhibit morphologies consistent with formation in an outflow.  

These three characteristics, variability, partial covering, and broad profiles are often found together in a single object, further supporting the idea that each individual characteristic likely indicates an outflow.  One well studied NAL outflow in J2123-0050 \citep{Hamann10} is a prime example of all three, exhibiting variability, partial covering, and broad profiles that still have FWHMs that are as narrow or narrower than many of the systems in J1023+5142.  

There is more tentative evidence for a quasar outflow origin in the apparent line--lock between the \civ\ doublets in systems 1 and 2.  Line--locking, where the difference in outflow velocities of two systems is exactly the velocity separation of the doublet, means that the lines are being radiatively accelerated directly towards the observer.  The reality of the line--locking in this case is unclear, due to the difference in derived metallicities between the two systems.  Nevertheless, the incredibly small velocity offset (\S~\ref{subsec:id}), along with the very small probability for chance alignments \citep{Ganguly03} suggest that the phenomenon may be real.  The possible line--lock in \civ\ in systems 1 and 2 suggests that they are both part of an outflow and that these weak \civ\ lines play a significant role in radiatively driving the flow.  If this is really the case here, the gas probably originated near the source of radiative acceleration, i.e. the quasar.

Finally, we note a trend in line width with velocity shift away from
the quasar.  The narrowest lines, with FWHM~$\sim 20$~\kms\ are
closest to the quasar redshift.  The lines appear progressively
broader as the velocity shift increases, with the broadest system
described as a (narrow) mini-BAL with FWHM~$= 270$~\kms, shown in
Figure~\ref{fig:civ}.  A similar phenomenon has been observed before in other quasars with multiple \civ\ absorption lines clearly forming in outflows \citep{Hamann97, Steidel90, Hamann10}.  Although it provides no direct information on the absorber locations, the appearance of this pattern in J1023+5142 supports the idea that at least some of the systems form in a quasar outflow.  The tight grouping of all nine of the systems also suggests a relationship between them.  \citet{Ganguly03} determine that the probability of six similarly grouped NALs in the quasar RX J1230.8+0115 all forming in intervening (uncorrelated) gas at similar velocity shifts is extremely small.  The similarities between those NALs and the NAL complex in J1023+1542 implies a similarly small probability for all nine NALs in this complex forming independently in intervening gas.  Although there could be up to several interlopers in the NAL complex of J1023+5142 that might form in nearby galaxies in the line of sight, the density of these galaxies required to form all of the absorbers in the complex is beyond any expectations of cluster density at this redshift.

Overall, we conclude that at least six out of the nine systems originate in a highly structured outflow driven by the quasar, because they exhibit one or more of the following properties: partial covering, broad profile shapes, large line strengths, tight grouping with other systems, and proximity to the quasar redshift.  Systems 5 and 6 are the most likely outflow candidates because they exhibit partial covering as well as several of the other properties listed above.  Systems 4 and 8 are likely outflows because of their strong, broad, asymmetric and smooth profiles and systems 7 and 9 are probably outflows because of their broad and smooth shapes.  Systems 1 through 3 are more ambiguous in origin, with narrow widths, complete covering, lower velocity shifts and smaller strengths.  However systems 1 and 2 exhibit line-locking, which could be evidence of an outflow.  Ultimately, we find strong evidence that systems 4, 5, 6, 7, 8 and 9 are part of a quasar outflow, whereas systems 1, 2, and 3 could consist of intervening gas from the IGM or other galaxies in the line of sight.

\subsection{Outflow Properties}
\label{sec:out}

As described in \S~\ref{sec:loc}, the evidence suggests that the majority of absorption lines in this grouping are part of a complex quasar outflow.  This flow must be highly structured, with at least 6 and as many as 9 distinct absorbing structures along the line of sight.  The velocities in the 6 most secure outflow systems range from -2120 to -4760~\kms.  Several of these systems (4, 6, 7 and 8) also have super--thermal line widths, indicative of large turbulence or strong radial velocity sheer across the outflow structure.  

Some of the outflow structures, represented by systems 5 and 6, must be spatially small to produce partial covering of the background emission source.  These lines lie on top of the very weak \civ\ broad emission line.  Therefore, nearly all ($>$80\%) of the flux beneath these lines is continuum emission and any partial covering below $C_f = 0.80$ can be ascribed to the continuum source and not the much larger BLR.  We estimate the diameter of the accretion disk continuum source at 1550~\AA\ to be $d \approx 0.03$~pc and the diameter of the \civ\ broad line region to be $d \approx 0.8$~pc, based on the scaling relations\footnote{We estimate the luminosity from the rest--frame flux at 1450~\AA\ measured in the SDSS spectrum, which, combined with the luminosity distance and the bolometric correction factor L=3.4*$\nu$L$_\nu$(1450), gives $\lambda$L$_\lambda$(1450\AA).  We measure the \civ\ emission line FHWM in the SDSS spectrum, and using Equation 7 in \citet{Vestergaard06}, derive a black hole mass of $\log M_{BH} = 9.8~M_\odot$.  Based on these values, we calculate an Eddington luminosity fraction of 0.8.  The black hole mass and Eddington luminosity fraction are then used in the scaling relations in \citet{Hamann08} to calculate the size of the continuum and broad line emission regions.} in \citet{Hamann08}.   To partially cover the emission source, the absorbing clouds should have characteristic sizes similar to or less than the BLR diameter, and possibly even less than the accretion disk diameter. 

If the absorbers are discrete clouds, their small sizes and substantial velocity dispersions should lead to fairly rapid dissipation in the absence of an external pressure (see \citet{Hamann08}).  In particular, the characteristic size of $d \sim 0.03$~pc or $d \sim 0.8$~pc combined with $b = 45$~\kms\ in the partial covering system 6 indicate a dissipation time of roughly $t_{dis} \sim d / b \sim 660$~yrs or $t_{dis} \sim 17,400$~yrs.  At the measured velocity of v~$= -3254$~\kms, this gas component would travel just $\sim 2$~pc or $\sim 60$~pc before dissipating.  A thorough discussion of the creation and survival of these absorbing structures is beyond the scope of this paper.  However, these simple arguments suggest that at least some of the outflow components we measure are very near their point of creation.  

It is useful to compare the basic properties of this NAL outflow to BALs. The ionisations in both types of outflows are similar, with very little low ionisation gas (e.g. \ciii).  The outflow velocities of the NALs in J1023+5142 are lower, $\leq6200$~\kms, than the typical outflow velocities in BALs, which can reach up to $\ge20,000$~\kms\ \citep{Korista93}, but they do overlap.  The NALs have total H column densities ($N(H) \leq 10^{17.2}$ to $10^{19.1}$~cm$^{-2}$, individual values listed in Table~\ref{tab:abs}), more than 1000 times lower than typical BAL H column densities ($N(H) \ge 10^{20}-10^{22}$~cm$^{-2}$, and probably higher).  By definition, these NALs also have FWHMs around 1000 times narrower than typical BALs, and much smaller \textsc{REW}s as well.  Nonetheless, NALs like this might be part of the same general outflow phenomenon as BALs, viewed at different angles \citep{Elvis00, Ganguly01}.  

This complex of weak NAL outflows appears to be dramatically different from typical BAL outflows, and constitutes a nearly unexplored part of the quasar outflow phenomenon, with a range of physical parameters and kinematics more complex and varied than previously thought.  It is well-known that NALs are a common feature of quasar spectra.  Previous surveys have found that 40\% of quasars have \civ\ NALs, in particular 25\% have strong \civ\ NALs within v~$> -5000$~\kms\ \citep{Vestergaard03}, 60\% have quasar-driven outflows in some form, either BALs, NALs, or something in-between \citep{Ganguly08, Paola08}, and including high-velocity outflows raises the percentage to 70\% \citep{Misawa07a}.  If the coverage fraction of these outflows is less than 100\%, which is likely, they could be ubiquitous in the near quasar environment, and could potentially play an influential role in the physical processes occurring therein. 

Finally, we would like to understand what role the NAL outflow in J1023+5142 might have in feedback to galaxy evolution.  The low speeds and small column densities, e.g., compared to BAL flows, suggest that its feedback contribution is negligible.  However, there are large uncertainties relating to the outflow location and geometry.  For one particular NAL outflow at a derived radial distance of $\sim5$~pc, \citet{Hamann10} estimate that the kinetic energy yield is several orders of magnitude smaller than that necessary to influence feedback.  At the opposite extreme, \citet{Moe09} argue that the feedback contribution is significant for another NAL outflow at a derived distance of $\sim2-5$~kpc.  The location of the NAL outflow in J1023+5142 is not known well enough to make these estimates.  A more sensitive search for variability in these NALs could be very helpful for refining both the location and the total energy yield (see \citet{Hamann10} and references therein).  

\subsection{Metallicity}
\label{sec:metal}

We find greater than or consistent with solar abundances in all of the absorption systems in J1023+5142 except in system 1, in agreement with previous studies of narrow associated absorption at lower redshifts \citep{Petitjean99, Hamann01, Dodorico04, Gabel06}.  These high metallicities are consistent with the results of other studies of intrinsic gas as well, including BEL gas \citep{Hamann02} and therefore consistent with our interpretation that the gas is intrinsic to the quasar.  Intervening absorbers generally have very low metallicities, with $Z$ no more than a few hundredths solar, although there are cases where high metallicity intervening gas has been observed \citep{Prochaska06, Schaye07}.  We argue that the high metallicities found in 8 of the 9 systems in this quasar are consistent with locations near the quasar, however, we do not rely solely on this argument to determine the gas location.  Instead, we consider that high metallicities could be a general phenomenon found in all gas in the quasar host environment \citep{Prochaska08}.  
 
The high metallicities of the NAL systems in J1023+5142 require that its host galaxy had vigorous star formation in the epoch before the quasar was observable, leading to metal-rich gas in the quasar outflows \citep{Falomo08}. This evidence, along with previous studies of BELs lead us to conclude that the generally accepted paradigm of quasar-host galaxy evolution is correct, where a major merger leads to a vigorous burst of star formation, which then funnels gas to the centre of the galaxy and ignites a quasar that eventually blows out obscuring gas and dust to become visibly luminous \citep{PerezGonzalez08, Hopkins08, RamosAlmeida09}.  However, larger samples are needed to examine the full range of NAL properties and study their relationships to quasar outflows and host galaxy environments.  Measurements at high redshifts are particularly valuable because this is the main epoch of host/massive galaxy formation when the NAL gas might have a close relationship to ongoing or recent star formation in the hosts.  

\section{Summary}
\label{sec:sum}

We use NALs to improve our understanding of the evolutionary relationship between the central black hole and its host galaxy through the study of their location, origin and abundance information in high redshift quasars.  Here, we examine the properties of nine NAL systems in the quasar J1023+5142 and find $N(H)$ from $\leq 10^{17.2}$ to $10^{19.1}$ cm$^{-2}$, velocities from --1400 to --6200~\kms, \civ\ doppler $b$ values from 7 to 150~\kms, \civ\ \textsc{REW} from 0.02 to 0.81~\AA\ and two systems with partial covering of either the continuum or the BLR, at the level of $C_f \approx 0.7$, which imply absorber diameters of $\leq 0.03$~pc or $\leq 0.8$~pc.  

The NAL systems all appear to be highly ionised; none of the systems exhibit low ionisation species such as Si~\textsc{II}, C~\textsc{II} or Si~\textsc{III}, whereas all contain \civ\ and some contain higher ionisation species such as \ovi\ and \nv.  

The \civ\ absorption NALs are tightly grouped, suggesting that they have a physical relationship to one-another, and the proximity of the NAL complex to the quasar redshift suggests that the physical relationship includes the quasar itself. The range in velocity across the complex is larger than can be easily explained by a single galaxy or even by a large cluster of galaxies.  A more likely explanation is that the NAL complex formed in a multi-component quasar-driven outflow. 

We estimate directly the location of each system in J1023+5142 through the use of several diagnostics and find strong evidence (partial covering, broad and smooth profiles compared to thermal widths, velocities greater than galaxy dispersion velocities, supersolar metallicities) that systems 4, 5, 6, 7, 8 and 9 are part of a quasar outflow.  Systems 1, 2, and 3 have more ambiguous origins because they exhibit narrow widths, lower velocity shifts, and system 1 has a lower metallicity, so these systems could consist of intervening gas from the IGM or other galaxies in the line of sight. 

Systems 2--9 in J1023+5142 exhibit supersolar metallicities ranging from $Z \ge 1$ to $\ge 8~Z_{\odot}$.  System 1 has a lower metallicity of $Z \leq 0.3~Z_{\odot}$. The high metallicities are consistent with scenarios of galaxy and black hole formation and evolution.  

The NALs in outflows appear to be part of a related outflow complex, which is very different than other known outflow regions such as BAL outflows, and constitutes a relatively unexplored part of the quasar outflow phenomenon.  The outflows in J1023+5142 could be important for feedback between the black hole and the host galaxy, depending on the radial distance of the gas from the quasar.  

The narrow widths of NALs mean that detailed studies of individual objects like this are the only way to make progress in understanding this type of outflow.  Variability studies could be useful to add more examples of NAL outflows to the current sample available for similar detailed analysis.  

We will add significantly to this sample in future work, including detailed studies of the full range of NAL properties in 24 quasars at high redshift, during the main epoch of host/massive galaxy formation.  

\section{Acknowledgments}
This work was supported in part by a NASA Chandra award (TM9-0005X) and a grant from the Space Telescope Science Institute (HST-GO-11705).


\begin{thebibliography}{}

\bibitem[\protect\citeauthoryear{{Aldcroft}, {Bechtold} \& {Foltz}}{{Aldcroft}
  et~al.}{1997}]{Aldcroft97}
{Aldcroft} T.,  {Bechtold} J.,    {Foltz} C.,  1997, in {Arav} N.,  {Shlosman}
  I.,   {Weymann} R.~J.,  eds, Mass Ejection from Active Galactic Nuclei
  Vol.~128 of Astronomical Society of the Pacific Conference Series,
  {Variability in Radio-Loud Quasar Associated Absorption Lines}.
pp 25--+

\bibitem[\protect\citeauthoryear{{Anderson}, {Weymann}, {Foltz} \& {Chaffee}
  Jr.}{{Anderson} et~al.}{1987}]{Anderson87}
{Anderson} S.~F.,  {Weymann} R.~J.,  {Foltz} C.~B.,    {Chaffee} Jr. F.~H.,
  1987, \aj, 94, 278

\bibitem[\protect\citeauthoryear{{Arav}, {de Kool}, {Korista}, {Crenshaw}, {van
  Breugel}, {Brotherton}, {Green}, {Pettini} \& {et al.}}{{Arav}
  et~al.}{2001}]{Arav01}
{Arav} N.,  {de Kool} M.,  {Korista} K.~T.,  {Crenshaw} D.~M.,  {van Breugel}
  W.,  {Brotherton} M.,  {Green} R.~F.,  {Pettini} M.,    {et al.} 2001, \apj,
  561, 118

\bibitem[\protect\citeauthoryear{{Arav}, {Kaastra}, {Kriss}, {Korista}, {Gabel}
  \& {Proga}}{{Arav} et~al.}{2005}]{Arav05}
{Arav} N.,  {Kaastra} J.,  {Kriss} G.~A.,  {Korista} K.~T.,  {Gabel} J.,
  {Proga} D.,  2005, \apj, 620, 665

\bibitem[\protect\citeauthoryear{{Arnaud} \& {Rothenflug}}{{Arnaud} \&
  {Rothenflug}}{1985}]{Arnaud85}
{Arnaud} M.,  {Rothenflug} R.,  1985, \aaps, 60, 425

\bibitem[\protect\citeauthoryear{{Barlow}, {Hamann} \& {Sargent}}{{Barlow}
  et~al.}{1997}]{Barlow97}
{Barlow} T.~A.,  {Hamann} F.,    {Sargent} W.~L.~W.,  1997, in {Arav} N.,
  {Shlosman} I.,   {Weymann} R.~J.,  eds, Mass Ejection from Active Galactic
  Nuclei Vol.~128 of Astronomical Society of the Pacific Conference Series,
  {Partial Coverage and Time Variability of Narrow-Line Intrinsic QSO
  Absorption Systems}.
pp 13--+

\bibitem[\protect\citeauthoryear{{Bergeron} \& {Herbert-Fort}}{{Bergeron} \&
  {Herbert-Fort}}{2005}]{Bergeron05}
{Bergeron} J.,  {Herbert-Fort} S.,  2005, ArXiv Astrophysics e-prints

\bibitem[\protect\citeauthoryear{{Braun} \& {Milgrom}}{{Braun} \&
  {Milgrom}}{1989}]{Braun89}
{Braun} E.,  {Milgrom} M.,  1989, \apj, 342, 100

\bibitem[\protect\citeauthoryear{{Dietrich}, {Hamann}, {Shields}, {Constantin},
  {Heidt}, {J{\"a}ger}, {Vestergaard} \& {Wagner}}{{Dietrich}
  et~al.}{2003}]{Dietrich03}
{Dietrich} M.,  {Hamann} F.,  {Shields} J.~C.,  {Constantin} A.,  {Heidt} J.,
  {J{\"a}ger} K.,  {Vestergaard} M.,    {Wagner} S.~J.,  2003, \apj, 589, 722

\bibitem[\protect\citeauthoryear{{D'Odorico}, {Cristiani}, {Romano}, {Granato}
  \& {Danese}}{{D'Odorico} et~al.}{2004}]{Dodorico04}
{D'Odorico} V.,  {Cristiani} S.,  {Romano} D.,  {Granato} G.~L.,    {Danese}
  L.,  2004, \mnras, 351, 976

\bibitem[\protect\citeauthoryear{{Elvis}}{{Elvis}}{2000}]{Elvis00}
{Elvis} M.,  2000, \apj, 545, 63

\bibitem[\protect\citeauthoryear{{Falomo}, {Treves}, {Kotilainen}, {Scarpa} \&
  {Uslenghi}}{{Falomo} et~al.}{2008}]{Falomo08}
{Falomo} R.,  {Treves} A.,  {Kotilainen} J.~K.,  {Scarpa} R.,    {Uslenghi} M.,
   2008, \apj, 673, 694

\bibitem[\protect\citeauthoryear{{Ferland}, {Korista}, {Verner}, {Ferguson},
  {Kingdon} \& {Verner}}{{Ferland} et~al.}{1998}]{Ferland98}
{Ferland} G.~J.,  {Korista} K.~T.,  {Verner} D.~A.,  {Ferguson} J.~W.,
  {Kingdon} J.~B.,    {Verner} E.~M.,  1998, \pasp, 110, 761

\bibitem[\protect\citeauthoryear{{Foltz}, {Weymann}, {Peterson}, {Sun},
  {Malkan} \& {Chaffee} Jr.}{{Foltz} et~al.}{1986}]{Foltz86}
{Foltz} C.~B.,  {Weymann} R.~J.,  {Peterson} B.~M.,  {Sun} L.,  {Malkan} M.~A.,
     {Chaffee} Jr. F.~H.,  1986, \apj, 307, 504

\bibitem[\protect\citeauthoryear{{Fox}, {Bergeron} \& {Petitjean}}{{Fox}
  et~al.}{2008}]{Fox08}
{Fox} A.~J.,  {Bergeron} J.,    {Petitjean} P.,  2008, ArXiv e-prints, 806

\bibitem[\protect\citeauthoryear{{Gabel}, {Arav} \& {Kim}}{{Gabel}
  et~al.}{2006}]{Gabel06}
{Gabel} J.~R.,  {Arav} N.,    {Kim} T.-S.,  2006, \apj, 646, 742

\bibitem[\protect\citeauthoryear{{Ganguly}, {Bond}, {Charlton}, {Eracleous},
  {Brandt} \& {Churchill}}{{Ganguly} et~al.}{2001}]{Ganguly01}
{Ganguly} R.,  {Bond} N.~A.,  {Charlton} J.~C.,  {Eracleous} M.,  {Brandt}
  W.~N.,    {Churchill} C.~W.,  2001, \apj, 549, 133

\bibitem[\protect\citeauthoryear{{Ganguly} \& {Brotherton}}{{Ganguly} \&
  {Brotherton}}{2008}]{Ganguly08}
{Ganguly} R.,  {Brotherton} M.~S.,  2008, \apj, 672, 102

\bibitem[\protect\citeauthoryear{{Ganguly}, {Eracleous}, {Charlton} \&
  {Churchill}}{{Ganguly} et~al.}{1999}]{Ganguly99}
{Ganguly} R.,  {Eracleous} M.,  {Charlton} J.~C.,    {Churchill} C.~W.,  1999,
  \aj, 117, 2594

\bibitem[\protect\citeauthoryear{{Ganguly}, {Masiero}, {Charlton} \&
  {Sembach}}{{Ganguly} et~al.}{2003}]{Ganguly03}
{Ganguly} R.,  {Masiero} J.,  {Charlton} J.~C.,    {Sembach} K.~R.,  2003,
  \apj, 598, 922

\bibitem[\protect\citeauthoryear{{Ganguly}, {Sembach}, {Tripp}, {Savage} \&
  {Wakker}}{{Ganguly} et~al.}{2006}]{Ganguly06}
{Ganguly} R.,  {Sembach} K.~R.,  {Tripp} T.~M.,  {Savage} B.~D.,    {Wakker}
  B.~P.,  2006, \apj, 645, 868

\bibitem[\protect\citeauthoryear{{Gibson}, {Brandt}, {Schneider} \&
  {Gallagher}}{{Gibson} et~al.}{2008}]{Gibson08}
{Gibson} R.~R.,  {Brandt} W.~N.,  {Schneider} D.~P.,    {Gallagher} S.~C.,
  2008, \apj, 675, 985

\bibitem[\protect\citeauthoryear{{Grevesse}, {Asplund} \& {Sauval}}{{Grevesse}
  et~al.}{2007}]{Grevesse07}
{Grevesse} N.,  {Asplund} M.,    {Sauval} A.~J.,  2007, Space Science Reviews,
  130, 105

\bibitem[\protect\citeauthoryear{{Hamann}, {Barlow}, {Beaver}, {Burbidge},
  {Cohen}, {Junkkarinen} \& {Lyons}}{{Hamann} et~al.}{1995}]{Hamann95}
{Hamann} F.,  {Barlow} T.~A.,  {Beaver} E.~A.,  {Burbidge} E.~M.,  {Cohen}
  R.~D.,  {Junkkarinen} V.,    {Lyons} R.,  1995, \apj, 443, 606

\bibitem[\protect\citeauthoryear{{Hamann}, {Barlow}, {Junkkarinen} \&
  {Burbidge}}{{Hamann} et~al.}{1997}]{Hamann97}
{Hamann} F.,  {Barlow} T.~A.,  {Junkkarinen} V.,    {Burbidge} E.~M.,  1997,
  \apj, 478, 80

\bibitem[\protect\citeauthoryear{{Hamann}, {Beaver}, {Cohen}, {Junkkarinen},
  {Lyons} \& {Burbidge}}{{Hamann} et~al.}{1997}]{Hamann97c}
{Hamann} F.,  {Beaver} E.~A.,  {Cohen} R.~D.,  {Junkkarinen} V.,  {Lyons}
  R.~W.,    {Burbidge} E.~M.,  1997, \apj, 488, 155

\bibitem[\protect\citeauthoryear{{Hamann} \& {Ferland}}{{Hamann} \&
  {Ferland}}{1999}]{Hamann99}
{Hamann} F.,  {Ferland} G.,  1999, \araa, 37, 487

\bibitem[\protect\citeauthoryear{{Hamann}, {Kanekar}, {Prochaska}, {Murphy},
  {Milutinovic}, {Ellison} \& {Ubacks}}{{Hamann} et~al.}{2010}]{Hamann10}
{Hamann} F.,  {Kanekar} N.,  {Prochaska} J.~X.,  {Murphy} M.~T.,  {Milutinovic}
  N.,  {Ellison} S.,    {Ubacks} W.,  2010, in prep.

\bibitem[\protect\citeauthoryear{{Hamann}, {Korista}, {Ferland}, {Warner} \&
  {Baldwin}}{{Hamann} et~al.}{2002}]{Hamann02}
{Hamann} F.,  {Korista} K.~T.,  {Ferland} G.~J.,  {Warner} C.,    {Baldwin} J.,
   2002, \apj, 564, 592

\bibitem[\protect\citeauthoryear{{Hamann} \& {Sabra}}{{Hamann} \&
  {Sabra}}{2004}]{Hamann04}
{Hamann} F.,  {Sabra} B.,  2004, in {Richards} G.~T.,  {Hall} P.~B.,  eds, AGN
  Physics with the Sloan Digital Sky Survey Vol.~311 of Astronomical Society of
  the Pacific Conference Series, {The Diverse Nature of Intrinsic Absorbers in
  AGNs}.
pp 203--+

\bibitem[\protect\citeauthoryear{{Hamann} \& {Simon}}{{Hamann} \&
  {Simon}}{2010}]{Hamann08}
{Hamann} F.,  {Simon} L.,  2010, in prep.

\bibitem[\protect\citeauthoryear{{Hamann}, {Barlow}, {Chaffee}, {Foltz} \&
  {Weymann}}{{Hamann} et~al.}{2001}]{Hamann01}
{Hamann} F.~W.,  {Barlow} T.~A.,  {Chaffee} F.~C.,  {Foltz} C.~B.,    {Weymann}
  R.~J.,  2001, \apj, 550, 142

\bibitem[\protect\citeauthoryear{{Hartquist} \& {Snijders}}{{Hartquist} \&
  {Snijders}}{1982}]{Hartquist82}
{Hartquist} T.~W.,  {Snijders} M.~A.~J.,  1982, \nat, 299, 783

\bibitem[\protect\citeauthoryear{{Heckman}, {Lehnert}, {Strickland} \&
  {Armus}}{{Heckman} et~al.}{2000}]{Heckman00}
{Heckman} T.~M.,  {Lehnert} M.~D.,  {Strickland} D.~K.,    {Armus} L.,  2000,
  \apjs, 129, 493

\bibitem[\protect\citeauthoryear{{Hopkins}, {Hernquist}, {Cox} \& {Kere{\v
  s}}}{{Hopkins} et~al.}{2008}]{Hopkins08}
{Hopkins} P.~F.,  {Hernquist} L.,  {Cox} T.~J.,    {Kere{\v s}} D.,  2008,
  \apjs, 175, 356

\bibitem[\protect\citeauthoryear{{Hutsem{\'e}kers}, {Hall} \&
  {Brinkmann}}{{Hutsem{\'e}kers} et~al.}{2004}]{Hutsemekers04}
{Hutsem{\'e}kers} D.,  {Hall} P.~B.,    {Brinkmann} J.,  2004, \aap, 415, 77

\bibitem[\protect\citeauthoryear{{Jiang}, {Fan}, {Vestergaard}, {Kurk},
  {Walter}, {Kelly} \& {Strauss}}{{Jiang} et~al.}{2007}]{Jiang07}
{Jiang} L.,  {Fan} X.,  {Vestergaard} M.,  {Kurk} J.~D.,  {Walter} F.,  {Kelly}
  B.~C.,    {Strauss} M.~A.,  2007, \aj, 134, 1150

\bibitem[\protect\citeauthoryear{{Juarez}, {Maiolino}, {Mujica}, {Pedani},
  {Marinoni}, {Nagao}, {Marconi} \& {Oliva}}{{Juarez} et~al.}{2009}]{Juarez09}
{Juarez} Y.,  {Maiolino} R.,  {Mujica} R.,  {Pedani} M.,  {Marinoni} S.,
  {Nagao} T.,  {Marconi} A.,    {Oliva} E.,  2009, \aap, 494, L25

\bibitem[\protect\citeauthoryear{{Korista}, {Voit}, {Morris} \&
  {Weymann}}{{Korista} et~al.}{1993}]{Korista93}
{Korista} K.~T.,  {Voit} G.~M.,  {Morris} S.~L.,    {Weymann} R.~J.,  1993,
  \apjs, 88, 357

\bibitem[\protect\citeauthoryear{{Kuraszkiewicz} \& {Green}}{{Kuraszkiewicz} \&
  {Green}}{2002}]{Kuraszkiewicz02}
{Kuraszkiewicz} J.~K.,  {Green} P.~J.,  2002, \apjl, 581, L77

\bibitem[\protect\citeauthoryear{{Misawa}, {Charlton}, {Eracleous}, {Ganguly},
  {Tytler}, {Kirkman}, {Suzuki} \& {Lubin}}{{Misawa} et~al.}{2007}]{Misawa07a}
{Misawa} T.,  {Charlton} J.~C.,  {Eracleous} M.,  {Ganguly} R.,  {Tytler} D.,
  {Kirkman} D.,  {Suzuki} N.,    {Lubin} D.,  2007, \apjs, 171, 1

\bibitem[\protect\citeauthoryear{{Misawa}, {Eracleous}, {Charlton} \&
  {Kashikawa}}{{Misawa} et~al.}{2007}]{Misawa07c}
{Misawa} T.,  {Eracleous} M.,  {Charlton} J.~C.,    {Kashikawa} N.,  2007,
  \apj, 660, 152

\bibitem[\protect\citeauthoryear{{Moe}, {Arav}, {Bautista} \& {Korista}}{{Moe}
  et~al.}{2009}]{Moe09}
{Moe} M.,  {Arav} N.,  {Bautista} M.~A.,    {Korista} K.~T.,  2009, \apj, 706,
  525

\bibitem[\protect\citeauthoryear{{Nagao}, {Marconi} \& {Maiolino}}{{Nagao}
  et~al.}{2006}]{Nagao06a}
{Nagao} T.,  {Marconi} A.,    {Maiolino} R.,  2006, \aap, 447, 157

\bibitem[\protect\citeauthoryear{{Narayanan}, {Hamann}, {Barlow}, {Burbidge},
  {Cohen}, {Junkkarinen} \& {Lyons}}{{Narayanan} et~al.}{2004}]{Narayanan04}
{Narayanan} D.,  {Hamann} F.,  {Barlow} T.,  {Burbidge} E.~M.,  {Cohen} R.~D.,
  {Junkkarinen} V.,    {Lyons} R.,  2004, \apj, 601, 715

\bibitem[\protect\citeauthoryear{{Nestor}, {Hamann} \& {Rodriguez
  Hidalgo}}{{Nestor} et~al.}{2008}]{Nestor08}
{Nestor} D.,  {Hamann} F.,    {Rodriguez Hidalgo} P.,  2008, \mnras, 386, 2055

\bibitem[\protect\citeauthoryear{{Pentericci}, {Fan}, {Rix}, {Strauss},
  {Narayanan}, {Richards}, {Schneider}, {Krolik}, {Heckman}, {Brinkmann},
  {Lamb} \& {Szokoly}}{{Pentericci} et~al.}{2002}]{Pentericci02}
{Pentericci} L.,  {Fan} X.,  {Rix} H.-W.,  {Strauss} M.~A.,  {Narayanan} V.~K.,
   {Richards} G.~T.,  {Schneider} D.~P.,  {Krolik} J.,  {Heckman} T.,
  {Brinkmann} J.,  {Lamb} D.~Q.,    {Szokoly} G.~P.,  2002, \aj, 123, 2151

\bibitem[\protect\citeauthoryear{{P{\'e}rez-Gonz{\'a}lez}, {Rieke}, {Villar},
  {Barro}, {Blaylock}, {Egami}, {Gallego}, {Gil de Paz}, {Pascual}, {Zamorano}
  \& {Donley}}{{P{\'e}rez-Gonz{\'a}lez} et~al.}{2008}]{PerezGonzalez08}
{P{\'e}rez-Gonz{\'a}lez} P.~G.,  {Rieke} G.~H.,  {Villar} V.,  {Barro} G.,
  {Blaylock} M.,  {Egami} E.,  {Gallego} J.,  {Gil de Paz} A.,  {Pascual} S.,
  {Zamorano} J.,    {Donley} J.~L.,  2008, \apj, 675, 234

\bibitem[\protect\citeauthoryear{{Petitjean} \& {Srianand}}{{Petitjean} \&
  {Srianand}}{1999}]{Petitjean99}
{Petitjean} P.,  {Srianand} R.,  1999, \aap, 345, 73

\bibitem[\protect\citeauthoryear{{Popesso} \& {Biviano}}{{Popesso} \&
  {Biviano}}{2006}]{Popesso06}
{Popesso} P.,  {Biviano} A.,  2006, \aap, 460, L23

\bibitem[\protect\citeauthoryear{{Prochaska} \& {Hennawi}}{{Prochaska} \&
  {Hennawi}}{2009}]{Prochaska08}
{Prochaska} J.~X.,  {Hennawi} J.~F.,  2009, \apj, 690, 1558

\bibitem[\protect\citeauthoryear{{Prochaska}, {O'Meara}, {Herbert-Fort},
  {Burles}, {Prochter} \& {Bernstein}}{{Prochaska} et~al.}{2006}]{Prochaska06}
{Prochaska} J.~X.,  {O'Meara} J.~M.,  {Herbert-Fort} S.,  {Burles} S.,
  {Prochter} G.~E.,    {Bernstein} R.~A.,  2006, \apjl, 648, L97

\bibitem[\protect\citeauthoryear{{Ramos Almeida}, {Rodr{\'{\i}}guez Espinosa},
  {Barro}, {Gallego} \& {P{\'e}rez-Gonz{\'a}lez}}{{Ramos Almeida}
  et~al.}{2009}]{RamosAlmeida09}
{Ramos Almeida} C.,  {Rodr{\'{\i}}guez Espinosa} J.~M.,  {Barro} G.,  {Gallego}
  J.,    {P{\'e}rez-Gonz{\'a}lez} P.~G.,  2009, \aj, 137, 179

\bibitem[\protect\citeauthoryear{{Richards}, {York}, {Yanny}, {Kollgaard},
  {Laurent-Muehleisen} \& {vanden Berk}}{{Richards} et~al.}{1999}]{Richards99}
{Richards} G.~T.,  {York} D.~G.,  {Yanny} B.,  {Kollgaard} R.~I.,
  {Laurent-Muehleisen} S.~A.,    {vanden Berk} D.~E.,  1999, \apj, 513, 576

\bibitem[\protect\citeauthoryear{{Rodr\'iguez Hidalgo}, {Hamann}, {Nestor} \&
  {Shields}}{{Rodr\'iguez Hidalgo} et~al.}{2010}]{Paola08}
{Rodr\'iguez Hidalgo} P.,  {Hamann} F.,  {Nestor} D.~B.,    {Shields} J.,
  2010, in prep.

\bibitem[\protect\citeauthoryear{{Ruiz}, {Crenshaw}, {Kraemer}, {Bower},
  {Gull}, {Hutchings}, {Kaiser} \& {Weistrop}}{{Ruiz} et~al.}{2001}]{Ruiz01}
{Ruiz} J.~R.,  {Crenshaw} D.~M.,  {Kraemer} S.~B.,  {Bower} G.~A.,  {Gull}
  T.~R.,  {Hutchings} J.~B.,  {Kaiser} M.~E.,    {Weistrop} D.,  2001, \aj,
  122, 2961

\bibitem[\protect\citeauthoryear{{Ruiz}, {Crenshaw}, {Kraemer}, {Bower},
  {Gull}, {Hutchings}, {Kaiser} \& {Weistrop}}{{Ruiz} et~al.}{2005}]{Ruiz05}
{Ruiz} J.~R.,  {Crenshaw} D.~M.,  {Kraemer} S.~B.,  {Bower} G.~A.,  {Gull}
  T.~R.,  {Hutchings} J.~B.,  {Kaiser} M.~E.,    {Weistrop} D.,  2005, \aj,
  129, 73

\bibitem[\protect\citeauthoryear{{Rupke}, {Veilleux} \& {Sanders}}{{Rupke}
  et~al.}{2005}]{Rupke05b}
{Rupke} D.~S.,  {Veilleux} S.,    {Sanders} D.~B.,  2005, \apjs, 160, 115

\bibitem[\protect\citeauthoryear{{Schaye}, {Carswell} \& {Kim}}{{Schaye}
  et~al.}{2007}]{Schaye07}
{Schaye} J.,  {Carswell} R.~F.,    {Kim} T.-S.,  2007, \mnras, 379, 1169

\bibitem[\protect\citeauthoryear{{Shen}, {Strauss}, {Oguri}, {Hennawi}, {Fan},
  {Richards}, {Hall}, {Gunn}, {Schneider}, {Szalay}, {Thakar}, {Vanden Berk},
  {Anderson}, {Bahcall}, {Connolly} \& {Knapp}}{{Shen} et~al.}{2007}]{Shen07}
{Shen} Y.,  {Strauss} M.~A.,  {Oguri} M.,  {Hennawi} J.~F.,  {Fan} X.,
  {Richards} G.~T.,  {Hall} P.~B.,  {Gunn} J.~E.,  {Schneider} D.~P.,  {Szalay}
  A.~S.,  {Thakar} A.~R.,  {Vanden Berk} D.~E.,  {Anderson} S.~F.,  {Bahcall}
  N.~A.,  {Connolly} A.~J.,    {Knapp} G.~R.,  2007, \aj, 133, 2222

\bibitem[\protect\citeauthoryear{{Simon} \& {Hamann}}{{Simon} \&
  {Hamann}}{2010}]{Simon10}
{Simon} L.~E.,  {Hamann} F.,  2010, submitted to MNRAS

\bibitem[\protect\citeauthoryear{{Srianand} \& {Petitjean}}{{Srianand} \&
  {Petitjean}}{2000}]{Srianand00}
{Srianand} R.,  {Petitjean} P.,  2000, \aap, 357, 414

\bibitem[\protect\citeauthoryear{{Steidel}}{{Steidel}}{1990}]{Steidel90}
{Steidel} C.~C.,  1990, \apjs, 72, 1

\bibitem[\protect\citeauthoryear{{Trump}, {Hall}, {Reichard}, {Richards},
  {Schneider}, {Vanden Berk}, {Knapp}, {Anderson}, {Fan}, {Brinkman},
  {Kleinman} \& {Nitta}}{{Trump} et~al.}{2006}]{Trump06}
{Trump} J.~R.,  {Hall} P.~B.,  {Reichard} T.~A.,  {Richards} G.~T.,
  {Schneider} D.~P.,  {Vanden Berk} D.~E.,  {Knapp} G.~R.,  {Anderson} S.~F.,
  {Fan} X.,  {Brinkman} J.,  {Kleinman} S.~J.,    {Nitta} A.,  2006, \apjs,
  165, 1

\bibitem[\protect\citeauthoryear{{Tytler} \& {Fan}}{{Tytler} \&
  {Fan}}{1992}]{Tytler92}
{Tytler} D.,  {Fan} X.-M.,  1992, \apjs, 79, 1

\bibitem[\protect\citeauthoryear{{Tzanavaris} \& {Carswell}}{{Tzanavaris} \&
  {Carswell}}{2003}]{Tzanavaris03}
{Tzanavaris} P.,  {Carswell} R.~F.,  2003, \mnras, 340, 937

\bibitem[\protect\citeauthoryear{{Veilleux}, {Cecil} \&
  {Bland-Hawthorn}}{{Veilleux} et~al.}{2005}]{Veilleux05}
{Veilleux} S.,  {Cecil} G.,    {Bland-Hawthorn} J.,  2005, \araa, 43, 769

\bibitem[\protect\citeauthoryear{{Vestergaard}}{{Vestergaard}}{2003}]{Vesterga%
ard03}
{Vestergaard} M.,  2003, \apj, 599, 116

\bibitem[\protect\citeauthoryear{{Vestergaard} \& {Peterson}}{{Vestergaard} \&
  {Peterson}}{2006}]{Vestergaard06}
{Vestergaard} M.,  {Peterson} B.~M.,  2006, \apj, 641, 689

\bibitem[\protect\citeauthoryear{{Warner}, {Hamann} \& {Dietrich}}{{Warner}
  et~al.}{2004}]{Warner04}
{Warner} C.,  {Hamann} F.,    {Dietrich} M.,  2004, \apj, 608, 136

\bibitem[\protect\citeauthoryear{{Weymann}, {Carswell} \& {Smith}}{{Weymann}
  et~al.}{1981}]{Weymann81}
{Weymann} R.~J.,  {Carswell} R.~F.,    {Smith} M.~G.,  1981, \araa, 19, 41

\bibitem[\protect\citeauthoryear{{Weymann}, {Williams}, {Peterson} \&
  {Turnshek}}{{Weymann} et~al.}{1979}]{Weymann79}
{Weymann} R.~J.,  {Williams} R.~E.,  {Peterson} B.~M.,    {Turnshek} D.~A.,
  1979, \apj, 234, 33

\bibitem[\protect\citeauthoryear{{Wild}, {Kauffmann}, {White}, {York},
  {Lehnert}, {Heckman}, {Hall}, {Khare}, {Lundgren}, {Schneider} \& {vanden
  Berk}}{{Wild} et~al.}{2008}]{Wild08}
{Wild} V.,  {Kauffmann} G.,  {White} S.,  {York} D.,  {Lehnert} M.,  {Heckman}
  T.,  {Hall} P.~B.,  {Khare} P.,  {Lundgren} B.,  {Schneider} D.~P.,
  {vanden Berk} D.,  2008, \mnras, 388, 227

\end{thebibliography}
\end{document}